%% file: flow_final_control.tex
\documentclass[a4paper,11pt]{article}
\pdfoutput=1
\usepackage{mathptmx}
\usepackage[T1]{fontenc}
\usepackage{amssymb,amsmath}
\usepackage{bbold}
\usepackage[dvipsnames]{xcolor}
\usepackage{graphicx,color}
\usepackage{jheppub} 

\input HLdefs

\title{\boldmath Relative entropy as Renormalization Monotone}
\DeclareMathOperator{\csch}{csch}

\author[a]{Nima Lashkari,}
\author[b]{Hong Liu,}
\author[b]{Srivatsan Rajagopal}

\affiliation[a]{Institute for Advanced Study, Einstein Drive, Princeton, NJ, 08540, USA}
\affiliation[b]{Center for Theoretical Physics, Massachusetts Institute of Technology\\
77 Massachusetts Avenue, Cambridge, MA 02139, USA}

\emailAdd{lashkari@ias.edu}
\emailAdd{hong\_liu@mit.edu}
\emailAdd{srivat91@mit.edu}

\newcommand{\be}{\begin{equation}}
\newcommand{\ee}{\end{equation}}
\newcommand{\bea}{\begin{eqnarray}}
\newcommand{\eea}{\end{eqnarray}}
\newcommand{\nn}{\nonumber}
\newcommand{\cH}{\mathcal{H}}
\newcommand{\la}{\langle}
\newcommand{\lb}{\left(}
\newcommand{\rb}{\right)}
\newcommand{\ra}{\rangle}

\newcommand{\al}{\alpha}
\newcommand{\ep}{\epsilon}
\newcommand{\p}{\partial}

\newcommand{\mO}{\mathcal{O}}

\newcommand{\qed}{\nobreak \ifvmode \relax \else
      \ifdim\lastskip<1.5em \hskip-\lastskip
      \hskip1.5em plus0em minus0.5em \fi \nobreak
      \vrule height0.75em width0.5em depth0.25em\fi}
\newcommand{\bde}{\delta}

\abstract{
We develop new techniques for studying the modular and the relative modular flows of general excited states. We show that the class of states obtained by acting on the vacuum (or any cyclic and separating state) with invertible operators from the algebra of a region is dense in the Hilbert space. 
This enables us to express the modular and the relative modular operators, as well as the relative entropies of generic excited states in terms of the vacuum modular operator and the operator that creates the state. 
In particular, the modular and the relative modular flows of {\it any} state can be expanded in terms of the modular flow of operators in vacuum. We illustrate the formalism with simple examples including states close to the vacuum, and coherent and squeezed states in generalized free field theory.}

\begin{document}
\title{Modular Flow of Excited States}
\maketitle

\section{Introduction}

Entanglement and  quantum information have played increasingly important roles 
in our understanding of quantum field theory (QFT), equilibrium and non-equilibrium dynamics 
of strongly correlated condensed matter systems, and quantum gravity. 
Results from operator algebras 
and techniques developed in algebraic approach to quantum field theory provide  powerful tools for organizing 
and obtaining quantum information properties of QFT and quantum statistical systems~(see e.g.~\cite{Witten:2018zxz} for a recent review).

In the standard textbook approach to quantum field theory, the central object is the Hilbert space of physical states. 
 Algebraic quantum field theory (AQFT) was developed in the 1960s as an alternative approach  
where one treats the algebra of local physical observables as the central object~\cite{Haag:1963dh,Haag:1992hx,Araki:1999ar}. 
According to the Stone-von Neumann uniqueness theorem the algebra of observables of a quantum system with a finite number of degrees of freedom has a unique irreducible Hilbert space representation up to unitary transformations. Thus, the Hilbert space approach and algebraic approach are equivalent. For a system with an infinite number of degrees of freedom, like a QFT, the algebra of observables allows infinitely many inequivalent representations in the Hilbert space. The algebraic approach provides a more intrinsic description than any of the irreducible representation of the observables. For example, in a system with spontaneous symmetry breaking or phase transitions, different macroscopic states correspond to inequivalent representations of Hilbert space, but all share the same algebra of observables.

Compared with the standard approach, the algebraic approach is significantly more mathematical and abstract, and as a result, it is harder to perform explicit calculations in model theories. However, deep and rich mathematical structures have been uncovered about quantum field theory using this approach. These structures provide powerful tools to study  the quantum information properties of states in QFT and quantum statistical systems. For instance, 

\ben

\item  The Reeh-Schlieder theorem says that bounded operators restricted to an arbitrary 
open set in spacetime are enough to generate the full vacuum sector of the Hilbert space. It indicates 
that generic finite energy states in QFT are not only entangled, but also are entangled at all scales. Entanglement entropy in QFT is a property of the local algebra of observables and not just the states.

\item The Hilbert space of a continuum quantum field theory does not have a tensor product structure.
More explicitly, consider an open region $U$ on a Cauchy slice 
$\Sig$ with its complement given by $U_c$, one can not factorize the Hilbert space $\sH$ into $\sH_U \otimes \sH_{U_c}$.
The standard measures of entanglement such as entanglement entropy and Renyi entropies defined using the reduced density matrices are, thus ill-defined. Nonetheless, the algebraic approach provides intrinsic definitions of 
those quantum information quantities which are well defined in the continuum limit and free of ultraviolet ambiguities. An important object is Araki's relative modular operator, which provides an algebraic definition of relative entropy~\cite{Araki:1975zu,Araki:1976zv},
a central quantity in quantum information \cite{vedral2002role}.

\item According to the Unruh effect, a uniformly accelerated observer in the vacuum state of QFT feels a thermal bath at a temperature proportional to its proper acceleration. The emergence of a thermal  bath has to do with the entanglement structure of the vacuum state and the fact that the accelerated observer has no access to the region of spacetime that is outside of its causal horizon. There is a generalization of this effect to observers restricted to a general spacetime region. The Tomita-Takesaki theory defines a self-adjoint operator called the {\it modular Hamiltonian} which generates a {\it modular} time evolution. A local observer whose clock ticks with modular time finds itself in a thermal bath. The modular Hamiltonian and the modular flow provide powerful mathematical tools for dealing with the entanglement structure of a QFT.

\item The algebraic approach introduces many operator inequalities (e.g. the positivity of the modular and the relative modular operator, and the half-sided modular inclusion inequality~\cite{Borchers:2000pv, Witten:2018zxz}) which should provide tight global constraints on quantum information properties of QFT. 
\een

So far only bits of this deep and rich mathematical structure have been used, but they have already yielded remarkable results on long-standing questions. For example, a quantum version of the null energy condition was conjectured and proved recently~\cite{Bousso:2015mna,Balakrishnan:2017bjg}. Modular operator and relative entropy were also used to provide a precise formulation of the Bekenstein bound~\cite{Casini:2008cr}. Recently, the monotonicity of the relative modular operator was used to derive new inequalities for correlation functions of QFT \cite{Lashkari:2018nsl}.
The modular operator of a spherical region in a conformal field theory has a local expression in terms of the stress tensor~\cite{Hislop:1981uh} which has become an important tool for studying the entanglement properties of states of conformal field theory as well as those of holographic theories~\cite{Casini:2011kv, faulkner2014gravitation, lashkari2016gravitational}. See also~\cite{Wall:2011hj,Papadodimas:2013jku,Harlow:2016vwg,Casini:2017vbe,Harlow:2018tng} for other applications.

In holography there is strong evidence that modular flows play an important 
role in reconstructing the corresponding bulk spacetime from the boundary quantum field theory~\cite{Rehren:1999jn,Jafferis:2015del,Faulkner:2017vdd,Faulkner:2018faa}. 
A better understanding of modular flows and relative modular flows should yield 
new insights into various aspects of bulk reconstruction, including the emergence of bulk causality and 
a better physical understanding of the entanglement wedge from the perspective of the boundary theory.

Despite their importance, our understanding of modular flows  in QFT is limited~(previous results include~\cite{Hislop:1981uh, Bisognano:1976za, Borchers:1998, Saffary:2006, Longo:2009mn, Casini:2009, Casini:2009sr, Brunetti:2010, Tedesco:2014eaz, Klich:2015, Cardy:2016,casini2017modular,Klich:2017}).
 In each case, where the modular operator has been found, significant insights have been obtained. 
Very little is known about modular operators and modular flows in excited states and for general regions. One reason is that the modular operators for  general states
in general regions are believed to be highly nonlocal and complicated. Nevertheless, it is of great interest to characterize this non-locality and hopefully extract universal features.

In this paper, we develop  new techniques for studying the modular flow of excited states, and the relative modular operator of general excited states. A key observation is that generic excited states 
can be obtained by acting on the vacuum (or any cyclic and separating state) with invertible operators from the algebra of a region. More precisely, we show that such states are dense in the Hilbert space.  This observation enables us to express, in a simple way, the modular and the relative modular 
operator, as well as relative entropies, of generic excited states in terms of the modular operator of the vacuum. 
Since the modular or relative modular flows are continuous (see section~\ref{sec3}),  we can obtain the modular flow of any state from those of a dense set of states. Modular, relative modular operators or their unbounded functions (such as logarithm) are not continuous, thus for these quantities we have access to generic states, but not all states in the Hilbert space.
A related observation is that the modular flow in an excited state can be obtained from that of  
 the vacuum via the so-called {\it unitary cocycle}, which in general is easier to construct and study 
than the modular operator of an excited state. Equivalently, one could also obtain the modular flow in an excited state using the relative modular flow. Previous discussions of modular operators for excited states include~\cite{lashkari2016modular, Sarosi:2017rsq}.

We illustrate the formalism using some simple examples, including states near the vacuum, and coherent and squeezed states 
in generalized free field theory.
To our knowledge, the relative modular operator in QFT has not been explicitly worked out in any examples, not even in free theories.

The plan of the paper is as follows. 
In section \ref{sec2}, we review the algebraic approach to general quantum systems and the Tomita-Takesaki modular theory.
Furthermore, in Appendix~\ref{app:a} we discuss the modular theory from the perspective of quantum information theory using tensor diagrams. 

In section~\ref{sec3}, we first prove two lemmas regarding the class of states which can be generated by acting by invertible operators on the vacuum (or any cyclic and separating state). Then, we obtain their modular and relative modular operators. 
We also consider the structure of Kubo-Mori Fisher information metric in this light.

In section \ref{sec4}, we consider the excited states that correspond to coherent and squeezed coherent states in generalized 
free field theory, and work out explicit expressions for the modular flow of local operators in these states.

We conclude in Sec.~\ref{sec:conc} with a brief discussion of future directions.

\section{Essential aspects of algebraic quantum field theory }\label{sec2}

We start by reviewing the essential aspects of the algebraic approach to quantum systems which will be 
relevant for our discussion in this paper (see~\cite{Haag:1992hx} for a textbook review).

\subsection{Algebraic setup} \label{sec:set}

Consider a quantum field theory with field operators $\{A_i\}$, where the index $i$ labels different operators (including all operators and not just ``fundamental'' ones). 
We will  be concerned with bounded operators $\{A_i\}$ and the $*$-algebra $\sA$ they form, i.e. they are closed under multiplications, and allow an adjoint operation.

A state $\psi$ is defined as a linear functional $\psi(A): \sA \to \CC$ on the algebra, which is positive, i.e.   $\psi(A^\da A) \geq 0$ for all $A \in \sA$, and normalized $\psi(\mathcal{I}) =1$ ($\mathcal{I}$ denotes the identity operator). 
Given a state $\psi$, the Gelfand--Naimark--Segal (GNS) construction builds a Hilbert space $\sH$ that carries a representation, $\pi_\psi$,
of the $*$-algebra $\sA$. In particular, there exists a vector $\ket{\Psi} \in \sH$ for which $\psi(A) = \vev{\Psi|A|\Psi}$ for all
 $A \in \sA$. The choice of the state and its corresponding Hilbert space is highly non-unique, and depends on the nature of the theory and the physical questions under consideration. 
 
 For a quantum field theory in Minkowski spacetime, we assume that there exists a unique Poicare invariant state $\om$.  
 The GNS Hilbert space associated with $\om$ is referred to as the vacuum sector, and coincides with the usual 
 definition of the Hilbert space in the standard approach (e.g. in Wightman axioms). The vector $\ket{\Om}\in\sH$ that corresponds to state $\om$ will be referred to as the vacuum.  
 Throughout this paper, we restrict to the vacuum sector of theories in Minkowski
 spacetime.

Now consider an open region $\sO$ in Minkowski spacetime. The operators which have their supports inside $\sO$ also 
form a $*$-algebra, which we denote as $\sA (\sO)$. 
In particular, self-adjoint elements of $\sA (\sO)$ can be interpreted as observables which can be measured in $\sO$. 
We will take the set of $\sA (\sO)$ (for all $\sO$) to have the following properties: 

\ben 

\item $\sA (\sO)$ is closed under the weak convergence limit. 
More explicitly, we say a sequence of operators $\{A_n \in \sA (\sO)\}$ with $n=1,2, \cdots$ converge to an operator $A$ if 
all the matrix elements converge, i.e.  $\vev{\psi| A_n |\phi} \to \vev{\psi |A|\phi}$ for any $\ket{\psi}, \ket{\phi} \in \sH$. 
A $*$-algebra which is closed under the weak convergence limit is a von Neumann algebra.

An important property of a von Neumann algebra $\sM$ is that it is equal to its double commutant, $\sM'' = \sM$, 
where $\sM'$ denotes the commutant of $\sM$, i.e. the set of bounded operators which commute with $\sM$. 
Therefore, we have $\sA (\sO) = \sA'' (\sO)$.

\item For $\sO_1 \subset \sO_2$ we have $\sA(\sO_1) \subset \sA(\sO_2)$. 

\item For a Poincare transformation $g$, denoting the unitary action on $\sA$ that coresponds to $g$ by $\al_g$, we have 
\be \label{tran}
\al_g \sA (\sO) = \sA (g \sO)  \ .
\ee

\item $\sA(\sO_1)$ commutes with $\sA (\sO_2)$ if $\sO_{1,2}$ are spacelike separated, 
\be \label{comm} 
[\sA (\sO_1), \sA(\sO_2)] = 0 \ .
\ee

\item Let $\hat \sO$ denotes the causal completion of $\sO$, then 
\be \label{casu}
\sA(\hat \sO) = \sA (\sO) \ ,
\ee
which requires that the dynamical laws be consistent with the causal structure. Equation~\eqref{casu} can be 
derived if one assumes the so-called Haag duality
\be \label{hd} 
\sA (\sO') = \sA' (\sO),
\ee
where $\sO'$ denotes  the causal complement of $\sO$.  However, it is known that the Haag duality is violated in 
some situations. \cite{Leyland:1978iv}

\een

A  striking statement regarding the local operator algebra $\sA(\sO)$ is the Reeh-Schlieder theorem which says that 
 for any open region $\sO$, the set of vectors $\sA (\sO) \ket{\Om}$ is dense in $\sH$. 
 This theorem has far reaching consequences. For example, it means that by performing a local operation on the Earth one could in principle create, say a basketball, in the Andromeda galaxy. One could do so no matter what the size of region $\sO$ is.
The theorem implies that the state $\ket{\Om}$ must be entangled at all scales. Thus one could take advantage of the long range correlations in $\ket{\Om}$ judiciously to choose operators in $\sO$ to achieve the desired outcome just like in an EPR experiment. Clearly, to take full advantage of the entanglement one has to apply operators from the algebra that are in general non-unitary.
 
 Mathematically, what the Reeh-Schlieder theorem says is that  the vacuum state $\ket{\Om}$ is cyclic and separating with respect to $\sA (\sO)$ for any open region $\sO$. 
 We say $\ket{\Om}$ is cyclic with respect to $\sA (\sO)$ if  $\sA (\sO) \ket{\Om}$ is dense in $\sH$. We say that $\ket{\Om}$ is separating with respect to $\sA (\sO)$, if exists no operator $A \in \sA (\sO)$ such that $A \ket{\Om} = 0$. 
Consider another open region $\tilde \sO$ which is causally disconnected from $\sO$. From the Reeh-Schlieder theorem $\sA (\tilde \sO) \ket{\Om}$ is also dense in $\sH$. Now suppose there exists an $A \in \sA (\sO)$ such that $A \ket{\Om} = 0$, then 
we also have $A \tilde A \ket{\Om} = \tilde A A \ket{\Om} =0$ for any $\tilde A \in \sA (\tilde O)$. Since $\sA (\tilde \sO) \ket{\Om}$ is dense in $\sH$ this can only happen if $A =0$.  Note that the theorem can also be generalized to a state with bounded energy, and thus any such state is cyclic and separating with respect to $\sA (\sO)$.

\subsection{Algebraic approach to entanglement: a toy ``spacetime''}

To exhibit the entanglement structure implied by the Reeh-Schlieder theorem for a quantum field theory, let us first  
consider a toy model to build up some intuition. 

Consider a ``spacetime'' whose spatial manifold consists of only two points: $L$ and $R$. The full Hilbert space of the system has a tensor 
product structure $\sH = \sH_L \otimes \sH_R$ with isomorphic $\sH_{L,R}$ for each point. For simplicity, we take them to 
have finite dimension $D$. 
The local algebra for point $L$ is thus $\sA_L = \sB (\sH_L) \otimes \sI$ and that for point $R$ is $\sA_R = \sI \otimes \sB (\sH_R)$, 
where $\sB (\sH_{L,R})$ denotes the set of bounded operators on Hilbert space $\sH_{L,R}$ and $\sI$ is the identity operator. 
 The operator algebra for the full system is $\sA = \sA_L \otimes \sA_R$.
 Clearly $[\sA_L, \sA_R] =0$ as in~\eqref{comm}, and $\sA_L = \sA_R'$. Since $\sA_{L,R}'' = \sA_{L,R}$ both $\sA_{L,R}$ are von Neumann algebras. 

Now, consider a cyclic and separating state $\ket{\Om}$ for $\sA_L$. Using the Schmidt decomposition of a bipartite state  one can readily see that such a state can be written as 
\bea
|\Om\ra= \sum_{a=1}^D \sqrt{\lambda_a}|a\ra_L |a \ra_R,  \qquad \lam_a > 0, \qquad  \sum_{a=1}^D \lambda_a=1 \ 
\eea
with $\ket{a}, a =1, \cdots D$ some basis for $\sH_{L,R}$. Assuming no vanishing Schmidt coefficient $\lambda_a$ is equivalent to assuming that the state is cyclic and separating.
The reduced density matrix $\rho_\Om^{R}$ is obtained by tracing over $\sH_{L}$:
\bea
\rho_\Om^L =\sum_{a=1}^D \lambda_a |a\ra_L\, {_L}\la a| , 
\eea 
and has full rank. The same holds for $\rho_\Om^R$. Thus, the cyclic and separating condition means that $\ket{\Om}$ is fully entangled between $\sH_L$ and $\sH_R$
with an entanglement entropy 
\be \label{vene0}
S_{\Om} = - {\rm Tr}_L \le(\rho_\Om^L \log \rho_\Om^L \ri)=  -  {\rm Tr}_R  \le(\rho_\Om^R \log \rho_\Om^R \ri)=  - \sum_a \lam_a \log \lam_a \ .
\ee

\subsubsection{Modular conjugation and modular operators} 

We now explore the entanglement structure of a cyclic and separating state $\ket{\Om}$ from the perspective of the algebras $\sA$ and $\sA_{L,R}$.  It is convenient to introduce an unnormalized maximally entangled vector 
 \bea\label{EOmega}
 |E_\Om \ra\equiv \sum_a |a\ra_L|a\ra_R\ 
 \eea
in terms of which we can write $\ket{\Om}$ as $|\Om\ra=((\rho_\Om^L)^{\ha} \otimes \mathcal{I} ) |E_\Om \ra$. Note that the definition of $E_\Om$ 
depends on the choice of $\ket{\Om}$ which selects the basis $\ket{a}$. 
 
Using $\ket{E_\Om}$ we can define an anti-linear operator $J_\Om$ on $\sH$ as follows: for  $\ket{\phi} = 
\phi_{ab} \ket{a}_L \ket{b}_R$
\bea\label{Janti}
J_\Om \ket{\phi} 
 \equiv \phi_{ab}^* \, {}_R\la b|E_\Om \ra{}_L\la a|E_\Om\ra= \phi_{ab}^* \ket{b}_L \ket{a}_R \ .
\eea
It can be readily checked that $J_\Om$ is anti-unitary, satisfies $J_\Om^2 = 1$, i.e.  $J_\Om=J_\Om^{-1}=J_\Om^\dagger$, and 
\be 
J_\Om \ket{\Om} = \ket{\Om} \ . 
\ee
An interesting property of $J_\Om$ is that it takes an operator $A \in \sA_R$  to an operator in $\sA_L$, 
\be\label{JAJ0}
J_\Om \sA_L  J_\Om = \sA_R = \sA_L'  \ 
\ee
and vice versa. More explicitly, taking $A = A_{ab} \ket{a}_L  \, {_L}\bra{b}  \in \sA_L$, we find 
\be \label{JAJ}
J_\Om A J_\Om = A_{ab}^* \ket{a}_R \, {_R} \bra{b} \in \sA_R  \ . 
\ee
We will refer to $J_\Om$ as the modular conjugation operator. 

Now, introduce the modular operator 
\be 
\De_\Om = \rho_\Om^L \otimes (\rho_\Om^R)^{-1} = e^{- K^L_\Om} \otimes e^{K^R_\Om}
\ee
with 
\be \label{denl}
K^R_\Om = - \log \rho_\Om^R , \qquad K^L_\Om = - \log \rho_\Om^L , \qquad K^L_\Om = J_\Om K^R_\Om J_\Om \ .
\ee
The operator $\De_\Om$  is positive and satisfies 
\be 
\De_\Om \ket{\Om} = \ket{\Om}, \qquad J_\Om \De_\Om J_\Om = \De_\Om^{-1} \ .
\ee
The modular operator $\De_\Om$ can be used to define a unitary flow for $\sA_{L,R}$ respectively,
\be 
U_\Om (s) \sA_L U_\Om^\da (s) = \sA_L, \quad  U_\Om (s) \sA_R U_\Om^\da (s) = \sA_R , \quad U_\Om (s) \equiv \De_\Om^{- i s} \ .
\ee
More explicitly, for $A \in \sA_R$ and $A' \in \sA_L$, we have 
\be 
U_\Om (s) A U_\Om^\da (s) = e^{i K_R s} A e^{-i K_R s} \in \sA_R, \quad U_\Om (s) A' U_\Om^\da (s) = e^{-i K_L s} A' e^{i K_L s} \in \sA_L \ .
\ee
Physically,  $U_\Om (s)$ defines a ``local'' time evolution under which an observer in $R$ (or $L$) remains 
in $R$ (or $L$).\footnote{In contrast, a generic Hamiltonian $H$ defined on $\sH$ will not preserve $\sA_{R,L}$ under evolution.}
In particular, under such an evolution, the $R$ (or $L$) observer  experiences a thermal state with
inverse temperature $\beta = 1$.

Finally, let us consider the anti-linear
\be 
S_\Om \equiv J_\Om \De^\ha_\Om
\ee
which has the properties
\be
S_\Om^2 = 1 , \quad S_\Om^\da S_\Om = \De_\Om , \quad S_\Om S_\Om^\da = \De_\Om^{-1} , \quad S_\Om \ket{\Om} = \ket{\Om} \ .
\ee
From the action of $J_\Om$ and $\De_\Om$, one also finds that 
\bega \label{DefS}
S_\Om A |\Om \ra=A^\dagger|\Om\ra ,  \quad 
S_\Om^\da A' |\Om\ra=A'^\dagger|\Om\ra , \quad   \forall A\in \mathcal{A}_R , \forall A' \in \mathcal{A}_L  \ .
\end{gather}

To summarize, the states $\ket{\Om}$ that are cyclic and separating in $\sA_{L,R}$  are entangled in every physical mode between $L,R$. 
On one hand, this can be seen from the fact that the corresponding reduced density matrices $\rho^{L,R}_\Om$ are full rank.  
On the other hand,  being cyclic and separating leads to the existence of $J_\Om, \De_\Om, S_\Om$, and their properties. Thus, the algebraic structure can be considered as an alternative way to probe the entangled 
nature of $\ket{\Om}$. This is of particular importance in quantum field theory where the density matrix is not well-defined, nonetheless the algebraic approach continues to hold. All the algebraic relations discussed above can be conveniently represented using tensor diagrams which make them more intuitive. See Appendix~\ref{app:a}.

\subsubsection{Relative modular operator and relative entropy} 

Now, consider a second state $\ket{\Psi}$. We can find the corresponding $\rho_\Psi^L$ by tracing over $\sH_R$. 
The relative entropy between $\rho_\Psi^L$ and $\rho_\Om^L$ can be written as 
\be \label{relen}
S(\Psi\|\Om) =  {\rm Tr}_L \le(\rho_\Psi^L \log \rho_\Psi^L - \rho_\Psi^L \log \rho_\Om^L \ri) \ .
\ee
Since $\rho_\Om^L$ is full rank the above quantity is well defined regardless of the nature of $\ket{\Psi}$. 

Introduce the relative modular operator between $\ket{\Psi}$ and $\ket{\Om}$ as 
\be \label{relm0}
\De_{\Psi \Om} =  \rho_\Psi^L\otimes (\rho^R_\Om)^{-1} \ .
\ee
The relative entropy~\eqref{relen} can be written in terms of $\De_{\Psi \Om}$ as
\be \label{rela}
S(\Psi\|\Om) = \vev{\Psi|\log \De_{\Psi\Om} |\Psi}  \ .
\ee

Suppose $\ket{\Psi}$ is also cyclic and separating with respect to $\sA_{L,R}$, i.e. 
\be 
\ket{\Psi} = \sum_a \sqrt{\sig_a} \ket{\tilde a}_L \ket{\tilde a}_R, \quad \sig_a > 0, \quad \sum_a \sig_a = 1 \ 
\ee
with $\{\ket{\tilde a}\}$ some other basis of $\sH_{L,R}$. 
We denote the modular conjugation, modular operator and the Tomita operator for the state $|\Psi\ra$, respectively, with $J_\Psi, \De_\Psi$ and $S_\Psi$.
If the basis  $\{\ket{\tilde a}\}$  coincides with $\{\ket{a}\}$ then $J_\Psi = J_\Om$, but in general this is not the case. 

We can introduce a relative conjugation operator $J_{\Psi \Om}$ as 
\be 
J_{\Psi \Om} \ket{v}_L \ket{w}_R = {_R}\vev{w|E_\Psi} {_L}\vev{v|E_\Om}
\ee
where $E_\Psi$ is defined analogously as $E_\Om$.  
One can check that $J_{\Psi \Om}$ is anti-unitary and satisfies 
\be 
J_{\Psi \Om}  J_{ \Om \Psi} = 1 , \quad J_{\Psi \Om} = J_{\Om \Psi}^\da \ .
\ee
With $\rho_\Psi$ full rank we can introduce its logarithm as in~\eqref{denl}, and write~\eqref{relm0} and the corresponding 
relative modular flow operator as 
\be \label{relm}
\De_{\Psi\Om}  
=  e^{- K_{\Psi}^L} e^{K_\Om^R} , \quad 
U_{\Psi \Om} (s) = \De_{\Psi \Om}^{-is} = e^{i K_{\Psi}^L s} e^{- i K_\Om^R s} \ .
\ee

We then find that 
\bega \label{relativemodflow}
U_{\Psi \Om} (s) A U_{\Psi \Om}^\da (s) = U_\Psi (s) A U^\da_\Psi (s), \quad A \in \sA_L \ , \\
  U_{\Psi \Om} (s) A' U_{\Psi \Om}^\da (s) = U_\Om (s) A' U^\da_\Om (s), \quad A' \in \sA_R \ .
\end{gather}
Also note that 
\be 
J_{\Psi \Om} \De_{\Psi \Om} J^\da_{\Psi \Om} = \De^{-1}_{\Om \Psi} \ .
\ee 
Introducing 
\be
S_{\Psi \Om} = J_{\Psi \Om} \De^\ha_{\Psi \Om}  
\ee
we find  
\be\label{relmoddef}
S_{\Psi \Om} A |\Om\ra = A^\da |\Psi\ra , \qquad S_{\Psi \Om}^\da  A' |\Om\ra = A'^\da |\Psi\ra  ,\quad \forall A \in \sA_L, \, \forall A' \in \sA_R \ . 
\ee

The unitary flow operator $U_{\Psi \Om}$ in~\eqref{relm} belongs to neither of $\sA_{L,R}$. We can 
also define unitary operators which belong to $\sA_{L,R}$, 
\bega  \label{uu3}
u_{\Psi \Om} (s)
= e^{ i s K_{\Psi}^L } e^{ - i s K_\Om^L}   =U_{\Psi \Om} (s) U^\da_\Om (s) \in \sA_L \\
u'_{\Psi \Om} (s)
= e^{- i s K_\Psi^R } e^{ i s K_\Om^R}   =U_{\Om \Psi} (s) U^\da_{ \Om} (s) \in \sA_R  \ .
\end{gather}

\subsection{Modular operator and modular flows in QFT}

Now, let us come back to quantum field theory. A key difference with the toy spacetime of previous subsection is that 
the Hilbert space of quantum field theory does not have a tensor product structure. More explicitly, consider an open region $U$ on a Cauchy slice 
$\Sig$ with its complement given by $U_c$, one can not factorize  the Hilbert space $\sH$ into $\sH_U \otimes \sH_{U_c}$.
In other words, the reduced density matrix associated with a region $U$ does not exist. 
Thus, we can no longer use~\eqref{vene0} and~\eqref{relen} to characterize the entanglement properties of 
a state $\ket{\Om}$ and the relative quantum information between $\ket{\Om}$ and $\ket{\Psi}$. 
 
Fortunately, thanks to the Tomita-Takesaki theory, even in the absence of tensor product structure and reduced density matrices, the algebraic structure 
discussed in previous subsection survives and can be used to capture entanglement properties of the system. 
 
Before stating the main results of the Tomita-Takesaki theory, we should note that it is common practice to put a quantum field theory on a lattice, where a tensor product structure for $\sH$ does exist, calculate the entanglement entropy~\eqref{vene0}, and then take the continuum limit. The continuum limit does not really exist as $\rho_\Om^{L,R}$ do not exist, which is reflected in that their corresponding entanglement entropies suffer from ultraviolet divergences and are sensitive to the short-distance cutoff. 
While it is often possible to extract valuable long-distance information from the divergent value,\footnote{An alternative is to use replica trick via Euclidean path integrals.}  it is clearly mathematically and physically preferable to directly deal with quantities which are intrinsically defined in the continuum.

Let $\sH$ be a Hilbert space and $\sM$ a von Neumann algebra acting on this space with $\sM'$ its commutant. 
Suppose the vector $\ket{\Om}$ is cyclic and separating for $\sM$. The Tomita-Takesaki theory asserts: 

\ben 
\item There exists an anti-unitary modular conjugation operator $J_\Om$ and a positive modular operator $\De_\Om$ satisfying 
\be 
J_\Om \ket{\Om} = \ket{\Om}, \quad \De_\Om \ket{\Om} = \ket{\Om} , \quad J_\Om=J_\Om^{-1}=J_\Om^\dagger, 
\quad  J_\Om \De_\Om J_\Om = \De_\Om^{-1} \ . 
\ee

\item $J_\Om$ takes an operator in $\sM$ to $\sM'$, i.e. 
\be 
J_\Om \sM  J_\Om = \sM'  \ .
\ee

\item $\De_\Om$ defines a unitary flow for $\sM, \sM'$ respectively,
\be 
U_\Om (s) \sM U_\Om^\da (s) = \sM, \quad  U_\Om (s) \sM' U_\Om^\da (s) = \sM' , \quad U_\Om (s) \equiv \De_\Om^{- i s} \ .
\ee

\item Let 
\be 
S_\Om \equiv J_\Om \De^\ha_\Om
\ee
then 
\be
S_\Om^2 = 1 , \quad S_\Om^\da S_\Om = \De_\Om , \quad S_\Om S_\Om^\da = \De_\Om^{-1} , \quad S_\Om \ket{\Om} = \ket{\Om} 
\ee
and 
\bega \label{DefS1}
S_\Om A |\Om \ra=A^\dagger|\Om\ra ,  \quad 
S_\Om^\da A' |\Om\ra=A'^\dagger|\Om\ra , \quad   \forall A\in \sM , \forall A' \in \sM'  \ .
\end{gather} 

\een
In the last subsection, we obtained these properties with the help of tensor product structure of the Hilbert space and 
the corresponding reduced density matrices. The Tomita-Takesaki theory tells us these structures are in fact direct consequences of being cyclic and separating even in the absence of a tensor product structure. 

As discussed in Sec.~\ref{sec:set}, the operator algebra $\sA  (\sO)$ for an open region $\sO$ is a von Neumann algebra and the vacuum state $\ket{\Om}$ is cyclic and separating with resect to $\sA (\sO)$ from the Reeh-Schlieder theorem, thus the properties listed above apply follow.

\subsection{Relative modular operator, relative entropy, and relative modular flows}

The Tomita-Takesaki theory can be generalized to give relative quantum information  between two states $\ket{\Psi}$ and $\ket{\Om}$
again only assuming that they are cyclic and separating:

\ben 

\item There exists an anti-unitary operator $J_{\Psi \Om}$ and a positive relative modular operator $\De_{\Psi \Om}$ which have the properties 
\be 
J_{\Psi \Om}  J_{ \Om \Psi} = 1 , \quad J_{\Psi \Om} = J_{\Om \Psi}^\da , \quad J_{\Psi \Om} \De_{\Psi \Om} J^\da_{\Psi \Om} = \De^{-1}_{\Om \Psi} \ .
\ee

\item The relative entropy can be obtained as 
\be \label{rela1}
S(\Psi\|\Om) =  -\vev{\Psi|\log \De_{ \Om\Psi} |\Psi}  \ .
\ee
Note that while both~\eqref{rela} and~\eqref{rela1} reduce to the expression in (\ref{relen}) for a system with a finite dimensional Hilbert space,
only~\eqref{rela1} applies to a quantum field theory. \footnote{For instance, if $\Psi=U|\Omega\ra$ for $U$ a unitary in the algebra, the expression (\ref{rela}) gives $S(\Psi\|\Om)=0$ which is clearly incorrect.} See also equation~\eqref{jek} below.

\item Introducing the relative modular flow 
\be \label{relm11}
U_{\Psi \Om} (s) = \De_{\Psi \Om}^{-is}  \ .
\ee
one obtains the unitary evolutions 
\bega 
U_{\Psi \Om} (s) A U_{\Psi \Om}^\da (s) = U_\Psi (s) A U^\da_\Psi (s), \quad A \in \sM  \ , \\
  U_{\Psi \Om} (s) A' U_{\Psi \Om}^\da (s) = U_\Om (s) A' U^\da_\Om (s), \quad A' \in \sM' \ .
\end{gather}

\item With 
\be
S_{\Psi \Om} = J_{\Psi \Om} \De^\ha_{\Psi \Om}  , \qquad S^\da_{\Psi \Om} = \De^\ha_{\Psi \Om} J^\da_{\Psi \Om} = J_{\Om \Psi }  \De^{-\ha}_{\Om \Psi }
\ee
then 
\be 
S_{\Psi \Om} A \Om = A^\da \Psi , \qquad S_{\Psi \Om}^\da  A' \Om = A'^\da \Psi  ,\quad \forall A \in \sM, \, \forall A' \in \sM' \ . 
\ee

\item One can also show that  \cite{connes1973classification}
\bega  \label{uu5}
u_{\Psi \Om} (s)
 \equiv U_{\Psi \Om} (s) U^\da_\Om (s) \in \sM \\
u'_{\Psi \Om} (s)
  \equiv U_{\Om \Psi} (s) U^\da_{ \Om} (s) \in \sM'  \ .
\end{gather}
Note that both $U_{\Psi \Om} (s)$ and $U^\da_\Om (s)$ are defined outside $\sM$ or $\sM'$. It is a highly nontrivial statement 
that the particular combinations above belong to $\sM$ and $\sM'$. $u_{\Psi \Om} (s)$, which is often referred as the unitary co-cycle, has the following properties (similarly with $u'_{\Psi \Om}$): 

\ben 

\item  Cocycle identity:
\be 
u_{\Psi \Om} (t_1 + t_2) =  u_{\Psi \Om} (t_1) \le( U_\Om (t_1) u_{\Psi \Om} (t_2)  U_\Om^\da (t_1) \ri)  \ .
\ee

\item Chain rule:
\be
 u_{\Psi \Om} (t) u_{\Om \Phi} (t) = u_{\Psi \Phi} (t) \ .
  \ee

 \item Intertwining property: 
\bega
 \label{ss1}
  u_{\Psi \Om} (t)  (U_\Om (t) A U_\Om^\da (t)) = (U_\Psi (t) A U_\Psi^\da (t)) u_{\Psi \Om} (t)  \\
  \label{ss2}
 u'_{\Psi \Om} (t)   (U_\Om (t) A' U_\Om^\da (t))   =  (U_\Psi (t) A' U_\Psi^\da (t)) u'_{\Psi \Om} (t)  
   \end{gather} 
\een
Given $ u_{\Psi \Om} (0) = 1$, we can introduce the {\it relative Hamiltonian} as 
\be \label{reH}
h_{\Psi \Om} = i {\p u_{\Psi \Om} (t)  \ov \p t} \biggr|_{t=0} \ . 
\ee
which can be written more explicitly as 
\be\label{jek}
h_{\Psi \Om} = K_{\Psi \Om} - K_\Om = K_\Psi - K_{\Om \Psi}  
\ee
with
\be 
K_\Om = - \log \De_\Om, \quad K_\Psi = - \log \De_\Psi, \quad  K_{\Om \Psi} = - \log \De_{\Om \Psi}, \quad 
K_{\Psi\Om} = - \log \De_{\Psi\Om} \ .
\ee
\een
We reiterate that even though we obtained the properties above earlier assuming a tensor product structure, the fact they continue to exist in the absence of the tensor product is a highly nontrivial mathematical statement that follows from the Tomita-Takesaki modular theory.

\section{Modular flows of excited states}\label{sec3}

In this section, we first develop a formalism to obtain the modular operator and the relative modular operator of general excited states
from the modular operator of the vacuum $\ket{\Om}$ (or any cyclic and separating state).  We then discuss these operators for near vacuum states in general theories, as an illustration of the formalism. In the next section, we apply the formalism to coherent and squeezed states in generalized free theories. 

\subsection{Dense sets of states} 

Our starting point is an open region $\sO$ in spacetime and its corresponding local von Neumann operator algebra $\sM = \sA (\sO)$. We denote the commutant by $\sM'$.  
The Reeh-Schlieder theorem tells us that the vacuum state $\ket{\Om}$ is cyclic and separating with respect to $\sM$. 

Connes and Stormer proved in ~\cite{connes1978homogeneity} that the set of excited states $UU'|\Omega\rangle$ with unitaries  $U\in\mathcal{M}$ and $U'\in\mathcal{M}'$ is {\it dense} in the Hilbert space.\footnote{The mathematical statement they showed is that a von Neumann algebra is a type $III_1$ factor (same type as the local algebra in QFT) if and only if the set of states $UU'|\Omega\rangle$ is total in the Hilbert space.} For a state created by local unitaries, it is straightforward to see that the anti-linear operator 
\bea
S_{UU'|VV'}=U' VS_\Om U^\dagger V'^\dagger
\eea
satisfies the equation the equation that defines the relative Tomita operator
\bea
S_{UU'|VV'} a |VV'\ra=a^\dagger |UU'\ra\ .
\eea
The relative modular operator corresponding to this state is then given by 
\bea
\Delta_{UU'|VV'}=V'U\Delta U^\dagger V'^\dagger\ .
\eea
Setting $V = U$ and $V' = U'$ in the above expression we obtain the Tomita operator and the modular operator of the excited state $UU'|\Omega\rangle$.

In many examples of interest, we may be interested in considering a state generated by an operator $\Psi \in \sM$, i.e. 
$\ket{\Psi} = \Psi \ket{\Om}$ (up to a normalization constant). From the Reeh-Schlieder theorem such states are dense in the Hilbert space, i.e. they come arbitrarily close to any state. Below, we consider a subset of 
such states which is also dense in the Hilbert space, and in the next subsection we obtain their modular operator and relative modular operator.  The states of interest are $\Psi|\Omega\rangle$ with $\Psi$ an invertible operator in the algebra. Note that an operator $\Psi$ is invertible if neither its point spectrum nor its residual spectrum contains a zero.

\smallskip

{\bf Lemma 1}: {\it  The vectors $\Psi|\Omega\ra$ and $\Psi^\dagger|\Omega\ra$ with $\Psi\in\mathcal{M}$ are separating if and only if $\Psi$ and $\Psi^\dagger$ are invertible. The inverse operators need not be bounded, but they are densely defined in the Hilbert space. Moreover, any such state is automatically cyclic. }

{\bf Proof:} 
First note that if $\Psi$ is invertible $\Psi^\dagger$ is also invertible. This is because every $\Psi\in\mathcal{M}$ admits a polar decomposition $\Psi= W |\Psi|$ with $W$ a partial isometry and $|\Psi|$ self adjoint. Invertible $\Psi$ necessarily implies that $W$ is a unitary and $|\Psi|$ is invertible, which in turn implies that $\Psi^\dagger$ must be invertible; see Appendix \ref{AppSrivatsan}. 

The converse statement is simple to prove. The vector $\Psi|\Omega\ra$ is not separating if there exists an $A \in \sM$ such that $A|\Psi\ra=0$. However, this implies that the operator $A\Psi\in\mathcal{M}$ kills $|\Om\ra$ which contradicts the separating property of $\ket{\Om}$.  In general, it is possible that $A\Psi=0$, however this is never the case when $\Psi$ is invertible, as $\Psi^{-1}$ is densely defined in the Hilbert space. This establishes the converse statement.

Now, suppose $\Psi|\Omega\ra$ is separating. The operator $\Psi^\dagger$ is invertible if there are no zeros neither in its point spectrum nor in its residual spectrum. 
By definition, the point spectrum of $\Psi^\dagger$ contains zero if there exists a vector $|\Psi_0\ra$ such that 
$\Psi^\dagger|\Psi_0\ra=0$. 
Since $\Psi|\Omega\ra$ is separating, there exists a sequence of operators $a'_n \in \sM'$ such that
$\lim_n a_n' \Psi \ket{\Om}$ gives $\ket{\Psi_0}$, i.e. 
\bea
\lim_n\la\Psi_0|a'_n\Psi|\Omega\ra=1 \ .
\eea
Since $[a'_n,\Psi]=0$ we find 
\bea
\lim_n\la\Psi_0|\Psi a'_n|\Omega\ra=1
\eea
which cannot be the case if $\Psi^\dagger$ has a zero eigenvector. We thus conclude $\Psi^\dagger$ cannot have a zero in its point spectrum. Since $\Psi^\dagger|\Omega\ra$ is also separating we conclude that both $\Psi$ are $\Psi^\dagger$ are injective.  The operator $\Psi$ is bounded, hence its action is defined on all vectors in the Hilbert space. Now consider any $a' \in \sM'$, then the set $\{a' \Psi \ket{\Om}\}$ is dense in the Hilbert space due to the separating property of $\Psi|\Omega\ra$. Since $\Psi a'|\Omega\ra=a'\Psi|\Omega\ra$ and $\ket{\Om}$ is separating, we conclude that the range of $\Psi$ is also dense. This means that there are no zeros in the residual spectrum of neither $\Psi$ nor $\Psi^\dagger$. 
To summarize, we have established that if $\Psi|\Om\ra$ and $\Psi^\dagger|\Omega\ra$ are separating both $\Psi$ and $\Psi^\dagger$ are invertible. If $\Psi$ has a zero in its continuous spectrum, then its inverse is unbounded.

Finally, we show that if $\Psi|\Omega\ra$ and $\Psi^\dagger|\Om\ra$ are separating they are also cyclic. Suppose they are not cyclic, then, there exists an $a'\neq0\in\mathcal{M'}$ such that $a'\Psi|\Omega\rangle=\Psi a'|\Omega\rangle=0$. Since the vacuum is cyclic $a' \ket{\Om} \neq 0$ and 
we then find a contradiction with the statement proved earlier that $\Psi$ does not have a zero in its point spectrum. 
This concludes the proof.

We now show that the set of states of Lemma 1 is, in fact, dense in the Hilbert space. 

{\bf Lemma 2: } Let $\mathcal{M}$ denote a von Neumann algebra acting on a Hilbert space $\mathcal{H}$. Let $|\Omega\rangle$ be a cyclic and separating vector.  Let $G(\mathcal{M})=\left\lbrace a\in \mathcal{M}: a^{-1}\in\mathcal{M}\right\rbrace$. 
Then, $G(\mathcal{M})|\Omega\rangle$ is dense in $\sH$. 

{\bf Proof :}  We first note a theorem of~\cite{dixmier:1971,robertson:1976} which says that 
$G(\sM)$ is a dense subset of $\mathcal{M}$ in the strong operator topology. 

Let $|\chi\rangle$ be an arbitrary vector in the Hilbert space. Then, from the cyclicity of $\ket{\Om}$, we can construct a sequence $a_n\in\mathcal{M}$ such that
\begin{align}
\text{lim}_n a_n|\Omega\rangle = |\chi\rangle
\end{align}
which means, given $\epsilon>0$, there is an $N_\epsilon$ such that
\begin{align}
|||\chi\rangle-a_n|\Omega\rangle||<\epsilon\hspace{20pt}\forall n\geq N_{\epsilon}\ .
\label{convergence1}
\end{align}

For each $n$, from the fact that $G(\mathcal{M})$ is strongly dense in $\mathcal{M}$, we can construct a sequence $b_{m,n}\in G(\mathcal{M})$ such that 
\begin{align}
\text{lim}_{m} b_{m,n}|\Omega\rangle = a_n|\Omega\rangle
\end{align}
which  means that there are $M_n$  such that
\begin{align}
|| b_{m,n}|\Omega\rangle-a_n|\Omega\rangle||<1/n\hspace{20pt}\forall m\geq M_n  \ .
 \label{convergence2}
\end{align}
Introduce a new sequence $b_n$ defined by
\begin{align}
b_n = b_{M_n+1,n}\ .
\end{align}
We would like to show that for any $\ep$ there exists a $K_\ep$ such that 
\be
|||\chi\rangle-b_n|\Omega\rangle|| \leq \ep, \quad \forall n > K_\ep \ .
\ee
For this purpose, we choose $\ep_1 = \ha \ep$ in~(\ref{convergence1}) and $K_\ep = {\rm max} (N_{\ep/2}, 2/\ep)$. Then for any $n > K_\ep$ we have  
\begin{align}
|||\chi\rangle-b_n|\Omega\rangle||\leq |||\chi\rangle-a_n|\Omega\rangle||+||a_n|\Omega\rangle-b_n|\Omega\rangle||< {\ep \ov 2} +\frac{1}{n} < {\ep \ov 2} +{ \ep \ov 2}  = \ep \ .
\end{align}
This concludes the proof. 

Finally, we would like to remark on the continuity properties of the relative modular operator. Consider two sequences of $|\Phi_n\ra$ and $|\Psi_n\ra$ that converge to $|\Phi\ra$ and $|\Psi\ra$, respectively, with $|\Psi\rangle$ cyclic separating. For any bounded function $f$ we have the continuity property \cite{Araki:1976zv}:
\bea
\lim_n f\lb\Delta_{\Phi^{C}_n\Psi^{C}_n}\rb=  f\lb\Delta_{\Phi^{C}\Psi}\rb
\eea
in the strong operator topology. Here, $|\Phi_n^C\rangle,|\Psi_n^C\rangle,|\Phi^C\rangle$ are the vector representatives of $|\Phi_n\rangle,|\Psi\rangle,|\Phi\rangle$ in the so-called natural positive cone of $|\Psi\rangle$.
Since the modular flow of any state is independent of the precise vector we choose to represent the state with, this implies we can obtain the modular flow of any state from those of the dense set of states created by invertible operators by taking the limit
\bea
\text{lim}_{n}\Delta_{\Phi_n}^{it}A\Delta_{\Phi_n}^{-it}=\lim_n(\Delta_{\Phi_n\Psi_n}^{i t} A \Delta_{\Phi_n\Psi_n}^{-i t}) =(\Delta_{\Phi\Psi}^{it}A\Delta_{\Phi\Psi}^{-it}) = \Delta_\Phi^{it}A\Delta_\Phi^{-it}\ .
\eea
again in the strong operator topology. This statement also holds true for non-cyclic states.

However, logarithm is an unbounded operator and as was argued in \cite{Araki:1976zv} relative entropy is only lower semi-continuous. That is to say
\bea
\underline{\lim}_n S(\Phi_n\|\Psi_n)\leq S(\Phi\|\Psi)\ .
\eea
where $\underline{\lim}$ is the limit inferior of a sequence. This means that while we can compute the relative entropy of a dense set of states we do not have access to the relative entropy of limit states.

\subsection{Modular and relative modular operators} \label{sec:Mu}

In the previous subsection, we demonstrated that the set of states generated by invertible $\Psi$ is dense in the Hilbert space. It can be readily seen that the Tomita operator $S_\Psi$ for such a state $|\Psi\ra$ is 
\bea
&&S_\Psi=(\Psi^\dagger)^{-1}S_\Omega \Psi^\dagger \ .
\eea 
Indeed, 
\bea
&&S_\Psi \: A \Psi|\Omega\ra=(\Psi^\dagger)^{-1}S_\Omega \Psi^\dagger A\Psi|\Omega\ra=(\Psi^\dagger)^{-1}\Psi^\dagger A^\dagger \Psi|\Omega\ra=A^\dagger \Psi|\Omega\ra, \quad \forall A \in \sA (\sO) \\
&&S_\Psi^{\da} \: A' \Psi|\Omega\ra = \Psi S_\Om^\da \Psi^{-1} A' \Psi \ket{\Om}  = \Psi S_\Om^\da A' \ket{\Om} 
=A'^\dagger \Psi|\Omega\ra\ , \quad \forall A' \in \sA' (\sO) \ 
\eea
where in the second line we have used the fact that $A'$ commutes with $\Psi$. 
The modular operator for $\ket{\Psi}$ can then be written as 
\be\label{genmodular}
\Delta_\Psi=S^\dagger_\Psi S_\Psi=\Psi S^\dagger_\Omega \Psi^{-1}(\Psi^{-1})^\dagger S_\Omega \Psi^\dagger 
  \ .
\ee
One can obtain the  modular conjugation operator $J_\Psi$ from $J_\Psi = S_\Psi \De_\Psi^{-\ha}$, which is somewhat complicated. 
From the definition of the positive part of an operator $|\Psi|^2=\Psi^\dagger\Psi$ we know that 
\eqref{genmodular} can be written as 
\be \label{mod1}
\Delta_\Psi=\Psi S_\Omega^\dagger (\Psi^\dagger\Psi)^{-1}S_\Omega\Psi^\dagger= \Psi \Delta_\Omega^{1/2} (\Psi_J^\dagger\Psi_J)^{-1}\Delta_\Omega^{1/2}\Psi^\dagger,
\ee
where $\Psi_J\equiv J_\Omega \Psi J_\Omega$ belongs to the commutant algebra, and we have used $J_\Omega=J_\Omega^\dagger=J_\Omega^{-1}$.
\footnote{Note that $(\Psi_J)^{-1}=(\Psi^{-1})_J$ and $(\Psi^\dagger)_J=(\Psi_J)^\dagger$. However,  $(|\Psi|^2)^{-1}\neq |\Psi^{-1}|^2$ and here we denote $(|\Psi|^2)^{-1}$ by $|\Psi|^{-2}$.}

Let us consider the relative modular operator between $\ket{\Psi}$ and another state $\ket{\Phi} = \Phi \ket{\Om}$ where $\Phi$ is not necessarily invertible. To simplify the notation let us assume that both $\Psi$ and $\Phi$ are normalized such that 
$\vev{\Psi|\Psi} = \vev{\Phi|\Phi} =1$. Then we have 
\be 
S_{\Phi \Psi } = (\Psi^\dagger)^{-1}S_\Omega\Phi^\dagger
\ee
which indeed satisfies 
\bega 
S_{\Phi \Psi } A \ket{\Psi} = (\Psi^\dagger)^{-1}S_\Omega\Phi^\dagger A \Psi \ket{\Om} =(\Psi^\dagger)^{-1}  \Psi^\da A^\da \Phi \ket{\Om} = A^\da \ket{\Phi}, \quad \forall A \in \sA (\sO) \ , \\
S_{\Phi \Psi }^\da A' \ket{\Psi} =\Phi S_\Omega^\da \Psi^{-1} A' \Psi \ket{\Om} =\Phi  A'^\da  \ket{\Om} = A'^\da \ket{\Phi}, \quad \forall A' \in \sA' (\sO) \ .
\end{gather}
The relative modular operator can be written as  
\be\label{relmodop2}
\Delta_{\Phi\Psi}=S_{\Phi\Psi}^\dagger S_{\Phi\Psi}= 
\Phi S_\Omega^\dagger (\Psi^{-1})(\Psi^{-1})^\dagger S_\Omega \Phi^\dagger 
=  \Delta^{\ha}_\Om \Phi_{-{i \ov 2}} |\Psi_J|^{-2} \le(\Phi_{-{i \ov 2}} \ri)^\dagger \Delta^{\ha}_\Om 
\ee
where we have introduced $\Phi_{-{i \ov 2}} \equiv \De^{-\ha}_\Om \Phi \De^\ha_\Om$. 
Again $J_{\Phi \Psi}$ can be obtained indirectly from $S_{\Phi \Psi}$ and $\De_{\Phi\Psi}$. 
In particular, the relative modular operator when one state is the vacuum is simpler
\bea \label{mod2}
&&S_{\Phi \Om} =  S_\Omega\Phi^\dagger, \qquad \De_{\Phi \Om} = \Phi \De_\Om \Phi^\da \\
&&S_{\Om\Psi}=(\Psi^\dagger)^{-1}S_\Omega,\qquad \De_{\Om\Psi}=\Delta_\Omega^{1/2}|\Psi_J|^{-2}\De_\Omega^{1/2}\ .
\label{mod3}
\eea

Now we summarise a couple of important observations of Eq.(\ref{relmodop2}). 

Note that for Eq.(\ref{relmodop2}) to be well defined, we need $\Psi$ to be invertible.

Next, for a generic operator $\Phi\in\mathcal{A}$, if the operator $\Phi_{-i/2}$ is unbounded, it can be shown to be non-closeable\cite{YT}. In our later applications to perturbation theory, it is essential that this operator and its adjoint be both densely defined; therefore, its closeablity is essential.

To ensure this, given a von Neumann algebra $\mathcal{A}$, we consider one of its very important self adjoint subalgebras $\mathcal{A}_T$, called the \textit{Tomita subalgebra} of $\mathcal{A}$. This satisfies the following very important properties : i) The modular flow of elements in $\mathcal{A}_T$ is \textit{entire analytic} and ii) $\mathcal{A}_T$ is SOT dense in $\mathcal{A}$.\footnote{We thank Y. Tanimoto for pointing out the importance of this class of operators for our applications. For further discussions, see \cite{Takesaki}}

Practically, what this means, is if $\Phi$ is bounded, and chosen to be in $\mathcal{A}_T$, then $\Phi_{-i/2}$ automatically becomes bounded. Secondly, it is sufficient to consider Eq.(\ref{relmodop2}) between those $|\Phi\rangle, |\Psi\rangle$ states such that $|\Psi\rangle$ is created by an invertible operator and $|\Phi\rangle$ is created by an operator in the Tomita algebra. Since both these classes of states are dense in the Hilbert space, we can use the continuity argument of the previous subsection to compute modular flows for all states. We discuss such operators in an appendix.

Finally, we can ask for explicit examples of operators in $\mathcal{A}_T$. A particularly nice set of such operators is given by the smearing of modular evolved operators against gaussians (or really, any function holomorphic in a suitable strip in the complex plane)\bea \int_{-\infty}^\infty dt e^{-rt^2}\Phi(t) \hspace{20pt} r>0\hspace{10pt}\Phi(t)=\Delta^{-it}\Phi\Delta^{it}.\eea

In the subsequent, whenever the expression $\Phi_{-i/2}$ is encountered, such a smeared operator should be kept in mind.

Now consider a cyclic and separating $\Phi \ket{\Om}$. If we are interested in the modular flow of operators that are either in $\mathcal{A}$ or the commutant $\mathcal{A}'$ using the identity (\ref{relativemodflow}) we can replace the modular flow with the relative modular flow with one state chosen to be the vacuum. Comparing~\eqref{mod2} and~\eqref{genmodular} we see that $U_{\Psi \Om} (s) = \De_{\Psi \Om}^{-is}$ is in general much easier to obtain that $U_\Psi (s) =\De_\Psi^{- is}$. Thus, we use ~\eqref{relativemodflow} as a calculational tool since {\it the relative modular flows in general provide
a much easier way to obtain modular flows of excited states.} This will be the approach we use in next section when performing explicit calculations. 

In the special case that $\Psi= U\in\mathcal{A}$ is a unitary operator, various expressions above simplify. 
In such a case the corresponding $\ket{U}$ may be considered as a ``local'' state as any observables lying in the causal complement region $\sO_c$ has the same expectation value as the vacuum, 
\be 
\vev{U|A'|U} = \vev{\Om |U^\da A' U |\Om} = \vev{\Om|A'|\Om}, \qquad A' \in \sA (\sO_c) 
\ee
where we have used that $U$ and $A'$ commute. 
The modular operator and the corresponding modular flow are simply
\be\label{moduni}
\Delta_U=U\Delta_\Omega U^\dagger, \quad \Delta^{-i s}_U=U\Delta_\Omega^{-is}U^\dagger, \quad 
J_U = U J_\Om U^\da \ .
\ee
Similarly 
\be 
S_{U \Om} = S_\Om U^\da, \quad \De_{U \Om} = \De_U , \quad J_{U \Om} = J_\Om U^\da, \quad \De_{\Om U} = \De_\Om  \ 
\ee
and the corresponding unitary co-cycle~\eqref{uu5} $u_{U \Om} = 1$. 
Note for an arbitrary $\Phi$ which is not necessarily separating nor cyclic we find 
\bea\label{unitaryrelmod}
\Delta_{\Phi U}=\Phi \Delta_\Omega\Phi^\dagger = \De_{\Phi \Om} 
\eea
which is in fact independent of $U$. Finally, if $\Phi$ is also a unitary operator we have 
\be \label{unitarymodop}
S_{VU}=U S_\Omega V^\dagger, \quad \Delta_{VU}=V\Delta_\Omega V^\dagger = \De_V , \quad J_{VU}=U J_\Om V^\dagger,
\quad u_{VU} = U_{V \Om} U_{U \Om}^\da = U_V U_U^\da \ .
\ee
Clearly $u_{VU}$ acts trivially on any operator in $\sA' (\sO)$. 

In general, to compute the relative modular flow for non-unitary states ~\eqref{relm11} we need to obtain
\be 
K_{\Phi\Psi} = - \log \De_{\Phi\Psi}
\ee 
in an explicit form.
In next subsection, we discuss a perturbative series that achieves this. 

\subsection{A perturbative series}\label{sec:perb}

Consider two operators $\tilde \De$ and $\De$ which are close, in the sense that $\tilde \De - \De$ can be expanded in some small parameter. We will obtain a perturbative series for $\log \tilde \De - \log \De$ in terms of $\tilde \De - \De$ and 
unitarity flows $\De^{it}$ generated by $\De$. 

It can be shown that  $\log \tilde \De$ can be written as \cite{LLR} 
\be \label{poi}
 - \log \tilde \De =- \log \De + \sum_{m=1}^\infty Q_m
\ee
where  
\bega \label{poi1}
Q_m = 2 \pi \, \text{lim}_{\epsilon_i\rightarrow 0}\int dt_1\int dt_2...\int dt_m \, F_{\epsilon_i}(t_1,t_2,\cdots,t_m) \, \delta (t_1) \cdots \delta (t_m) \\
\delta (t) =\Delta^{-it}  \delta \Delta^{it}  ,  \quad \delta=\frac{\al}{1-\al/2}, \quad 
\al = 1-\Delta^{-\ha} \tilde \Delta\Delta^{-\ha}
\end{gather}
and the kernel $F$ is defined by
\begin{align}\label{poi2}
F_{\epsilon_i}(t_1,t_2,..,t_m)&=f(t_1)g_{\epsilon_1}(t_2-t_1)g_{\epsilon_2}(t_3-t_2)...g_{\epsilon_{m-1}}(t_{m}-t_{m-1})f(t_m) 
\end{align}
\be
f(t)=\frac{1}{2\text{cosh}(\pi t)}, \qquad 
g_{\epsilon}(t) =\frac{i}{4}\biggl[\frac{1}{\text{sinh}(\pi(t-i\epsilon))}+\frac{1}{\text{sinh}(\pi(t+i\epsilon))}\biggr] \ .
\ee
In~\eqref{poi1} it should be understood that the limit $\ep_i \to 0$ are taken after doing the integrals. 
A variant of~\eqref{poi}--\eqref{poi1} has appeared previously in~\cite{Sarosi:2017rsq}.\footnote{The expression in~\cite{Sarosi:2017rsq} is similar to~\eqref{poi1}, but there are some important differences: $\delta$ was not introduced there and the $i \ep$ prescription, the integration contours, and operator orderings also appear different.} 
The kernel $F_{\ep_i} (t_1,t_2,\cdots,t_m)$ has remarkable symmetric properties 
and one can write $Q_m$ in terms of nested commutators of $\bde$'s plus a set of contact terms~\cite{LLR}
\begin{align} \label{kiu}
Q_m 
& = {2 \pi \ov m} \text{lim}_{\epsilon_i\rightarrow 0}\int dt_1\int dt_2...\int dt_m \, F_{\epsilon_i}(t_1,t_2,...,t_m)[\cdots [\delta (t_1),\delta(t_2)], \cdots], \delta(t_m)] + P_m 
\end{align}
where $P_m$ are ``contact terms'' with only $m-2$ time integrations. For a complete description of contact terms see \cite{LLR}. Their structure for general $m$ is somewhat complicated which we will not need here. Below we only give the first few terms:
\bega
Q_1 = {\pi \ov 2} \int_{-\infty}^\infty  \frac{dt}{\cosh^2(\pi t)}  \de (t) , \\
Q_2 =  {\pi \ov 4} \int \frac{dt_1 dt_2 \, g_\ep (t_2-t_1) }{\cosh(\pi t_1)\cosh(\pi t_2)}\left[ \delta (t_1),  \de (t_2) \right]  , \\
Q_3 =  {\pi \ov 6} \int \frac{dt_1 dt_2dt_3 \, g_{\ep_1} (t_2-t_1)g_{\ep_2} (t_3-t_2)}{\cosh(\pi t_1)\cosh(\pi t_3)} [[\delta (t_1), \de (t_2)],\de(t_3) ] + \frac{\pi}{24}\int {dt \ov \cosh^2 (\pi t)} \, \delta^3 (t) , \\
Q_4 =  \frac{\pi}{8} \int 
\frac{dt_1 dt_2dt_3 dt_4 \, g_{\ep_1} (t_2-t_1)g_{\ep_2} (t_3-t_2) g_{\ep_3} (t_4-t_3)}{\cosh(\pi t_1)\cosh(\pi t_4)}
[[[\delta(t_1),\delta(t_2)],\delta(t_3)],\delta(t_4)] \cr
+ \frac{\pi}{32} \int \frac{dt_1 dt_2 \, g_\ep (t_2-t_1) }{\cosh(\pi t_1)\cosh(\pi t_2)} \left\lbrace\delta(t_2),[\delta(t_1),\delta^2 (t_2)]\right\rbrace
\end{gather}
where it should be understood that $\ep_i$'s  are taken to zero at the end.

\subsection{Modular and relative modular Hamiltonians for states close to vacuum}

Let us now apply the discussion of previous subsection to $K_{\Phi \Psi} = - \log  \De_{\Phi\Psi}$ 
with both $\ket{\Psi}$ and $\ket{\Phi}$ chosen to be close to the vacuum. From~\eqref{relmodop2} we find the corresponding $\al$ and $\de$ 
\be\label{alphadef}
\al = 1 -   \Phi_{-{i \ov 2}} |\Psi_J|^{-2} \le(\Phi_{-{i \ov 2}} \ri)^\dagger , \qquad
 \de = 2 {1- \Phi_{-{i \ov 2}} |\Psi_J|^{-2} \le(\Phi_{-{i \ov 2}} \ri)^\dagger\ov 1 + \Phi_{-{i \ov 2}} |\Psi_J|^{-2} \le(\Phi_{-{i \ov 2}} \ri)^\dagger} 
\ee
We consider  Hermitian operators $\Phi(\lambda)$ and $\Psi(\mu)$ which are close to the identity   
\be \label{pers}
\Phi (\lam) = 1 + \lam \phi^{(1)} + {\lam^2 \ov 2} \phi^{(2)} + \cdots , \qquad
\Psi (\mu) = 1 + \mu \psi^{(1)} + {\mu^2 \ov 2} \psi^{(2)} + \cdots \ 
\ee
with $\lam$ and $\mu$ some small continuous parameters. 
Normalizations for these states imply that 
\be \label{nor}
\vev{\Om|\phi^{(1)} |\Om} = \vev{\Om|\psi^{(1)}|\Om}  =  \vev{\Om|(\phi^{(1)})^2 + \phi^{(2)}|\Om}  =  \vev{\Om|(\psi^{(1)})^2 + \psi^{(2)} |\Om} =  0 \ .
\ee

We can now expand $\de$ and $K_{\Phi \Psi}$ in $\lam$ and $\mu$ as (with $K_\Om = - \log \De_\Om$)
\bea 
&& \de = \lam \de^{(10)} +\mu \de^{(01)} +\mu\lam\de^{(11)}+ {\lam^2 \ov 2} \de^{(20)} + {\mu^2 \ov 2} \de^{(02)} +\cdots ,\nn\\
&&K_{\Phi \Psi} - K_\Om =  \lam K^{(10)} +\mu K^{(01)} +\mu\lam K^{(11)}+ {\lam^2 \ov 2}K^{(20)} + {\mu^2 \ov 2} K^{(02)} +\cdots ,
\eea
where 
\bega \label{p1}
  \de^{(10)}= - \phi^{(1)}_{-{i \ov 2}}-\phi^{(1)}_{{i \ov 2}},\;\;   \de^{(01)}= 2\psi_J^{(1)},\;\; \de^{(11)}=0, \\
   \de^{(20)}=2\lb (\phi^{(1)})^2\rb_{-{i\ov 2}}+2\lb (\phi^{(1)})^2\rb_{{i\ov 2}},\;\;  \de^{(02)}=-4\lb\psi^{(1)}\rb_J^2, 
   \label{p2}
 \end{gather}
 and 
 \bega
K^{(10)} = {\pi \ov 2} \int_{-\infty}^\infty  \frac{dt}{\cosh^2(\pi t)} \de^{(10)} (t)  , \quad K^{(01)} = {\pi \ov 2} \int_{-\infty}^\infty  \frac{dt}{\cosh^2(\pi t)} \de^{(01)} (t)  , \quad
K^{(11)} = 0, \\
K^{(20)} = {\pi \ov 2} \int_{-\infty}^\infty  \frac{dt}{\cosh^2(\pi t)} \de^{(20)} (t)
 + {\pi \ov 2} \int \frac{dt_0 dt_1 \, g(t_1-t_0)}{\cosh(\pi t_0)\cosh(\pi t_1)}\left[ \de^{(10)} (t_0) ,   \de^{(10)} (t_1)  \right] , \\
 K^{(02)} = {\pi \ov 2} \int_{-\infty}^\infty  \frac{dt}{\cosh^2(\pi t)} \de^{(02)} (t)
 + {\pi \ov 2} \int \frac{dt_0 dt_1 \, g(t_1-t_0)}{\cosh(\pi t_0)\cosh(\pi t_1)}\left[ \de^{(01)} (t_0) ,   \de^{(01)} (t_1)  \right]  \ .
\end{gather}
Setting $\Phi$ and $\Psi$ in the above expressions to $ \Om$ or $\Psi$ or $\Phi$ we obtain, respectively, 
$K_{\Om \Phi}, K_{\Phi \Om}, K_\Psi$ and $K_\Phi$.

In~\eqref{p1}--\eqref{p2} we have used the following two operations for an operator $A\in\mathcal{M}$: 
\bea\label{spsis}
A_J=J_\Omega A J_\Omega\in\mathcal{M}',\qquad A_{\pm i/2}=\Delta_\Omega^{\pm 1/2}A\Delta^{\mp 1/2}_\Omega
\eea
where the operator $A_{i/2}$ is the modular flow of $A$ by an imaginary amount $-i/2$. 
The operators $A_{\pm i/2}$  commute with everything in $\mathcal{M}'$.
To see this, note that $A_{i \ov 2} = \De^\ha A\De^{-\ha} =J S AS J$ (suppressing $\Om$ subscript for $J$ and $\De$), and 
for any $B , C \in \sM$ 
\be 
[S A S, B] \ket{C} = (S A S B C - B S AS C )\ket{\Om} = (B C A^\da - B C A^\da) \ket{\Om} =0   \ .
\ee
Since $\ket{C}$ form a dense subset we conclude that 
\be
 [S A S, B] =0  \ .
\ee
Now for any operator $A' \in \sM'$ we can write it as $A' = J B J$ for some $B \in \sM$, we then conclude that 
\be \label{afi}
[A_{i /2} , A' ] = [J SAS J, J BJ] = J [SAS, B] J = 0  \ . 
\ee
Similarly for $A_{-i/2}$. 

Going back to~\eqref{p1}--\eqref{p2}. From~\eqref{afi} we have 
\be
[  \psi^{(1)}_{-{i \ov 2}} , \psi_J^{(1)}] = 0, \qquad [  \phi^{(1)}_{-{i \ov 2}} (t), \psi^{(1)}_J (t')] = 0
\ee
which can be used to obtain the relative Hamiltonian
\bea\label{idenK}
&&h_{\Phi \Psi } = K_{\Phi \Psi} - K_\Psi = K_\Phi - K_{\Psi \Phi}  \ . 
\eea
At the first order in $\lam$ and $\mu$ we have
\bea
&&K_{\Phi \Psi} - K_\Psi=\lam (K^{(10)}_{\Phi\Psi}-K^{(10)}_{\Psi\Psi}) +\mu (K^{(01)}_{\Phi\Psi}-K^{(01)}_{\Psi\Psi})+\cdots\nn\\
&&= {\pi \ov 2} \int_{-\infty}^\infty  \frac{dt}{\cosh^2(\pi t)}\lb \mu\lb \Psi_{-\frac{i}{2}}+\Psi_{\frac{i}{2}}\rb-\lam\lb \Phi_{-\frac{i}{2}}+\Phi_{\frac{i}{2}}\rb\rb +\cdots\nn\\
&&=\lam (K^{(10)}_{\Phi\Phi}-K^{(10)}_{\Psi\Phi}) +\mu (K^{(01)}_{\Phi\Phi}-K^{(01)}_{\Psi\Phi})+\cdots\ .
\eea
It can be checked explicitly that at higher orders in perturbation theory the identity (\ref{idenK}) is satisfied.

\subsection{Relative entropy and Fisher information} 

We now examine the behavior of relative entropy~\eqref{rela1} between two states generated by~\eqref{pers}.
We  expand $S(\Psi \|\Phi)$ explicitly in terms of $\lam,\mu$ as 
\bea
S(\Psi \|\Phi)=-\la \Psi|\log \Delta_{\Phi\Psi}|\Psi\ra
= S^{(0)} + \lam S^{(10)}+ \mu S^{(01)} +\lam\mu S^{(11)}+ {\lam^2 \ov 2} S^{(20)}  + {\mu^2 \ov 2} S^{(02)}  \cdots\nn
\eea
It can be readily seen that  $S^{(0)} = S^{(10)}=S^{(01)}  = 0$ due to the fact that
 $K_\Om \ket{\Om} = 0$, $\De_\Om\ket{\Om} = \ket{\Om}$, and  $\vev{\Om|\de^{(10)}_{\Phi \Psi} (t)|\Om} = \vev{\Om|\de^{(01)}_{\Phi \Psi}|\Om} = 0$ (from~\eqref{nor}). At order $\lam \mu$ we find the Kubo-Mori Fisher information 
 \bea
 F_\Om(\Psi,\Phi)&=&S^{(11)} (\Psi\|\Phi)= \vev{\Om\le|\{\psi^{(1)} , K_{\Phi\Psi}^{(10)}  \}\ri|\Om} \nn\\
 &=&-\frac{\pi}{2}\int\frac{dt}{\cosh^2(\pi t)}\vev{(\psi^{(1)})(\Delta^{1/2}+\Delta^{-1/2})\phi^{(1)}(t)}+h.c.\
 \eea
 which is manifestly symmetric in $\Psi$ and $\Phi$. Similar expressions for the Kubo-Mori Fisher information have been obtained in \cite{lashkari2016modular,Sarosi:2017rsq}
 
Furthermore, $S^{(20)}$ and $S^{(02)}$ are given by 
\bea
&&S^{(02)} (\Psi\|\Phi)=\vev{\Om\le|2\psi^{(1)} K_\Om \psi^{(1)} +2\{K_{\Phi \Psi}^{(01)},\psi^{(1)}\}+  K_{\Phi \Psi}^{(02)} \ri|\Om} \nn\\
&&S^{(20)} (\Psi\|\Phi)=\vev{\Om\le| K_{\Phi \Psi}^{(20)} \ri|\Om}\ .
\eea
Various terms in the above expressions can be written more explicitly as  
\bea
&&2\vev{\Om\le|\{K_{\Phi \Psi}^{(01)},\psi^{(1)}\}\ri|\Om}=4\pi\int\frac{dt}{\cosh^2(\pi t)}\vev{\Om|\psi^{(1)}\psi_J^{(1)}(t)|\Om}, \\
&&\vev{\Om\le|K_{\Phi \Psi}^{(02)}\ri|\Om}=-4\vev{\Om\le| (\psi^{(1)})^2\ri|\Om}+2\pi\int\frac{dt_0dt_1 g(t_1-t_0)}{\cosh(\pi t_0)\cosh(\pi t_1)}\vev{[\psi_J(t_0),\psi_J(t_1)]}\nn\\
&&=-4 \vev{\Om\le|(\psi^{(1)})^2 \ri| \Om}+\pi\int \frac{ds\:s g(s)}{\sinh(\pi s)}\vev{\Om|\psi^{(1)}(\Delta^{-i s}-\Delta^{i s})\psi^{(1)}|\Om} ,
\eea
where we have changed variables $s=t_1-t_0$ and $t=t_1+t_0$ in the last line, and used $J\Delta^{i s}J=\Delta^{i s}$. 
Similarly,
\be
\vev{\Om\le| K_{\Phi \Psi}^{(20)} \ri|\Om}=4 \vev{\Om\le|(\phi^{(1)})^2 \ri| \Om}+\pi\int \frac{ds\:s g(s)}{\sinh(\pi s)}\vev{\Om|\phi^{(1)}(\Delta^{1/2}+\Delta^{-1/2})^2(\Delta^{-i s}-\Delta^{i s})\phi^{(1)}|\Om} \ .
\ee

\section{Generalized free fields}\label{sec4}

As an illustration of the general formalism of Sec.~\ref{sec3}, in this section, we consider the modular and relative modular operators for coherent and squeezed states in a theory of generalized free fields.

Suppose $\phi$ is a generalized free field with commutator
\be \label{gcom}
[\phi(x),\phi(y)]=-i\Delta(x-y) ,
\ee
where $\Delta(x-y)$ is some distribution multiplied by the identity operator. For a region $\sO$, we consider smear fields $\Phi(f)=\int_\sO f(x) \phi(x)$ with 
$f(x)$ to have support in $\sO$ (i.e. vanish at the boundary of the region). 
The algebra of local operators in $\sO$ is generated by unitary operators $W(f)=e^{i\Phi(f)}$, with multiplication rule 
$W(f) W(g)=e^{i \sigma(f,g)} W(f+g)$,\footnote{In the case of a free scalar field we mod out $W(f)$ by the solutions to the Klein-Gordon equation of motion to make $\sigma(f,g)$ non-degenerate. In this case, $\sigma(f,g)$ is a symplectic form in phase space.} where $\sigma(f,g)$ is an anti-symmetric bilinear determined from~\eqref{gcom}.
We will take $\sO$ to be the right Rindler wedge $x^1>|t|$, for which the vacuum modular Hamiltonian $K_\Om$ is simply the boost operator, which acts on $\Phi (f)$ as 
\be\label{ft}
e^{i K_\Om t}\Phi(f)e^{-i K_\Om t}=\Phi(e^{t\mathcal{D}}f)\equiv\Phi(f_t), \qquad
\mathcal{D}=(x^1\p_0+x^0\p_1)\ .
\ee

\subsection{Unitary states}

\subsubsection{Coherent states} 

Consider the unitary coherent state
\be\label{co1}
|f\ra=U(f)|\Omega\ra, \qquad U(f)=e^{i\Phi(f)}\ .
\ee
Recall from our discussion of Sec.~\ref{sec:Mu} that the modular Hamiltonian $K_U$ for a state $\ket{U} = U \ket{\Om}$ generated from a unitary operator $U$ is 
\be \label{k1}
K_U=U K_\Omega U^\dagger \ .
\ee
As a result, the modular flow in these excited states act as
\bea \label{k2}
&&\forall a\in\mathcal{A} (\sO):\qquad e^{i K_U t}a e^{-i K_U t}=U e^{i K_\Omega t}(U^\dagger a U)e^{-i K_\Omega t}U^\dagger=Ad_{U}\lb(Ad_{U^{-1}}(a))(t)\rb \\
&&\forall a'\in\mathcal{A}(\sO)' :\qquad e^{i K_U t}a' e^{-i K_U t}=U e^{i K_\Omega t} a' e^{-i K_\Omega t}U^\dagger=Ad_U \lb a'(t)\rb
\label{k3}
\eea
where $Ad_U(a)=U a U^\dagger$ and $a(t)=e^{i K_\Omega t}ae^{-i K_\Omega t}$. Below for a self-adjoint operator $\Phi$ we will also use the notation $ad_\Phi a =[\Phi, a]$. We then have for~\eqref{co1}
\be
\Delta_f= e^{-2\pi K_f}\qquad K_f=U(f)K_\Omega U^\dagger(f)\ .
\ee

To compute~\eqref{k1}--\eqref{k3} we note the commutators 
\bea\label{Deltaanti-def}
&&[\Phi(f),\Phi(g)]=-i \int f(x)\Delta(x-y)g(y) \equiv -i \la f,\Delta g\ra \\
&&ad_{\Phi(f)}K_\Om=[\Phi(f),K_\Om]=\int_R f(x)[\phi(x),K_\Om]=-i\int_R f(x)\mathcal{D}\phi(x)=i\Phi(\mathcal{D}f) \\
&&ad_{\Phi(f)}^2K_\Om=[\Phi(f),[\Phi(f),K_\Om]]=\la f,\Delta\mathcal{D}f\ra \\
&&ad_{\Phi(f)}^nK_\Om=[\Phi(f),ad_{\Phi(f)}^{n-1}K_\Om]=0\qquad \forall n>2,
\eea
where we have used an integration by part since $f(x)$ vanishes on the boundary of $R$: $\mathcal{D}\Phi(f)=-\Phi(\mathcal{D}f)$. 
The series expansion that corresponds to $K_f$ truncates because the second nested operator is proportional to identity:
\bea\label{nested2}
K_f&=&\sum_{n=0}^\infty \frac{1}{n!}ad_{i\Phi(f)}^nK=K+[i\Phi(f),K]+\frac{1}{2}[i\Phi(f),[i\Phi(f),K]]+\cdots\nn\\
&=&K-\Phi(\mathcal{D}f)-\frac{1}{2}\la f,\Delta\mathcal{D}f\ra\ .
\eea
This form of the modular Hamiltonian is non-local; however, its non-locality is of the form of an integral over the Rindler wedge of a local operator, and a term that is bi-local in the region and proportional to the identity operator. Terms that are proportional to the identity operator do not contribute to the modular flow. To compute the modular evolution of an operator $\Phi(g)$ we need
\bea
&&ad_{K_f}\Phi(g)\equiv [K_f,\Phi(g)]=(-i)\lb \Phi(\mathcal{D}g)+\la f,\Delta \mathcal{D}g\ra\rb\nn\\
&&ad_{K_f}^2\Phi(g)\equiv [K_f,[ K_f,\Phi(g)]]=(-i)^2\lb \Phi(\mathcal{D}^2g)+\la f,\Delta \mathcal{D}^2g\ra\rb\nn\\
&&ad_{K_f}^n\Phi(g)\equiv [ K_f,\cdots [ K_f,\Phi(g)]]=(-i)^n\lb \Phi(\mathcal{D}^ng)+\la f,\Delta \mathcal{D}^ng\ra\rb\ .
\eea
The modular flow becomes
\be\label{modularflowunit}
e^{i t K_f}\Phi(g)e^{-i t K_f}=\sum_{n=0}^\infty \frac{1}{n!}ad_{i t K_f}^n\Phi(g)=\Phi(g_{2\pi t})+\la f,\Delta g_{2\pi t}\ra -\la f, \Delta g\ra, \quad
g_t=e^{t\mathcal{D}}g\ .
\ee

The relative modular flow of the state $|f\ra$ with respect to the vacuum is the same as the modular flow of the state $|f\ra$
\bea
\Delta_{f\Omega} = \Delta_f 
\eea
and the unitary cocyle for the state $|f\ra$ with respect to the vacuum $|\Omega\ra$ is
\bea
u_{f \Om} (t)=e^{ i t K_f}e^{-i t K_\Om} \ .
\eea

The above discussion does not depend on the nature of generalized free field $\phi$ or the specific form of the modular $K_\Om$. 
As far as $K_\Om$ and its modular flow are known we can obtain those of $K_f$. 
For $\phi$ an ordinary free scalar field with mass $m$, one can proceed alternatively. In this $K_\Om$ can be considered 
as an operator built from $\phi$ and its canonical momentum $\pi$, i.e $K_\Om = K_\Om (\phi, \pi)$. 
Note that 
\bea
&&Ad_U\phi(x)=U(f)\phi(x)U^\dagger(f)=\phi(x)+[i \Phi(f),\phi(x)]=\phi(x)+\tilde{f}(x),  \\
&&Ad_U\pi(x)=U(f)\pi(x)U^\dagger(f)=\phi(x)+[i\Phi(f),\pi(x)]=\pi(x)+\p_t\tilde{f}(x)
\eea
where
\bea
\tilde{f}(x)=\int_R f(y)\Delta(y-x)\equiv \la f,\Delta \delta\ra \ .
\eea
We can then write $K_f$ as 
\bea\label{modularHam}
K_f&=&K_\Om(\phi+\tilde{f},\pi+\p_t\tilde{f})\ .
\eea
Accordingly, the modular flow in this excited state can be written as 
\bea\label{modflowex}
&&e^{i K_f t}\phi(x)e^{-i K_f t}=U(f) e^{i K_\Om t}U(f)^\dagger \phi(x)U(f)e^{-i K_\Om t}U(f)^\dagger\nn\\
&&=U(f)\lb \phi(x_{2\pi t})-\tilde{f}(x)\rb U(f)^\dagger= \phi(x_{2\pi t})+\tilde{f}(x_{2\pi t})-\tilde{f}(x)\ .
\eea
Equation~\eqref{modflowex} is consistent with~\eqref{modularflowunit} if we choose $g = \delta (x-y)$ such that $\Phi (g) = \phi (x)$.

\subsubsection{Unitary squeezed states}

Now consider the state $|f_s\ra=e^{i\Phi(f)^2}|\Omega\ra$ where $\Phi(f)=\int_R f(x)\phi(x)$. These correspond to unitary squeezed coherent states because the operator $\Phi(f)^2$ is second order in creation and annihilation operators. 
Note that $\Phi(f)^2$ is bi-local. The modular Hamiltonian of this state is
\bea\label{Kf2}
K_{f_s}=e^{i\Phi(f)^2}K e^{-i\Phi(f)^2}
\eea
which is found by considering 
\bea
&&U(f_s)\phi(x)U^\dagger(f_s)=\phi(x)+i[\Phi(f)^2,\phi(x)]=\phi(x)+2\Phi(f)\tilde{f}(x)\nn\\
&&U(f_s)\pi(x)U^\dagger(f_s)=\pi(x)+i[\Phi(f)^2,\pi(x)]=\pi(x)+2\Phi(f)\p_t\tilde{f}(x)\nn\\
&&\tilde{f}(y)=\int_R f(x)\Delta(x-y)=\la f,\Delta \delta\ra\ .
\eea
Therefore, the modular Hamiltonian of $|f_s\ra$ and the relative modular Hamiltonian of $|f_s\ra$ with respect to the vacuum are
\bea
K_f=K_{f\Omega}=K\lb\phi(x)+2\Phi(f)\tilde{f},\pi(x)+2\Phi(f)\p_t\tilde{f}(x)\rb\ .
\eea
The modular flow of operator $\Phi(g)$ in this excited state is
\bea
e^{i t K_{f_s}}\Phi(g)e^{-i t K_{f_s}}&=&e^{i\Phi(f)^2}e^{i t K}e^{-i\Phi(f)^2}\Phi(g)e^{i\Phi(f)^2}e^{-i t K}e^{-i\Phi(f)^2}\nn\\
&=&e^{i\Phi(f)^2}e^{i t K}\lb \Phi(g)-2\Phi(f)\la f,\Delta g\ra\rb e^{-i t K} e^{-i\Phi(f)^2}\nn\\
&=&e^{i\Phi(f)^2}\lb\Phi(g_t)-2\Phi(f_t)\la f,\Delta g\ra \rb e^{-i\Phi(f)^2}\nn\\
&=&\Phi(g_t)-2\Phi(f_t)\la f,\Delta g\ra+2\Phi(f)\la f,\Delta g_t\ra-4\Phi(f)\la f_,\Delta f_t\ra\la f,\Delta g\ra 
\eea

The unitary cocyle flow is
\bea
&&(D f_2:D\Omega)(t)\phi(x)(D f_2:D\Omega)^\dagger(t)=\phi(x)+2\Phi(f)\tilde{f}(x)-2\tilde{f}(x_{-2\pi t})\Phi_{2\pi t}(f)\nn\\
&&\Phi_{t}(f)\equiv\int_R f(x)\phi(x_{t})+2\Phi(f)\int_R f(x)\tilde{f}(x_{t})\ .
\eea

Similar to the discussion of coherent states the BCH expansion can be used to obtain an expression for the modular Hamiltonian in terms of an integral over the Rindler wedge. To this aim, we compute the following commutators
\bea
&&ad_{\Phi(f)^2}K=[\Phi(f)^2,K]=i\lb \Phi(f)\Phi(\mathcal{D}f)+\Phi(\mathcal{D}f)\Phi(f)\rb\nn\\
&&[\Phi(f)^2,\Phi(g)\Phi(h)]=-2i\lb\Phi(f)\Phi(h)\la f,\Delta g\ra+\Phi(g)\Phi(f)\la f,\Delta h\ra\rb
\eea
where we have assumed that $f(x)$ vanishes on the boundary of the right Rindler wedge. Note that by anti-symmetry of $\Delta(x-y)$ we have $\la f,\Delta f\ra=0$. Therefore,
\bea
&&ad_{\Phi(f)^2}^2K=[\Phi(f)^2,[\Phi(f)^2,K]]=4\la f,\Delta\mathcal{D}f\ra \Phi^2(f)\nn\\
&&ad_{\Phi(f)^2}^nK=[\Phi(f)^2,\cdots,[\Phi(f)^2,[\Phi(f)^2,K]]]=0\qquad n>2\ .
\eea
Plugging this in the BCH expansion of \ref{Kf2} we find
\bea\label{BCHunitary}
&&K_{f_s}=K-\lb \Phi(f)\Phi(\mathcal{D}f)+\Phi(\mathcal{D}f)\Phi(f)\rb -2\la f,\Delta \mathcal{D}f\ra\Phi(f)^2,
\eea 
where we have used integration by parts $\mathcal{D}\Phi(f)=-\Phi(\mathcal{D}f)$.

\subsection{Non-unitary states}

Now, we consider non-unitary states generated by non-unitary operators of the type $e^{-\alpha\Phi(f)}$ and $e^{-\alpha\Phi(f)^2}$. 
These states are of physical interest as they may be considered as representing states generated by unitary operators
whose supports lie outside of the region $\sO$. For instance, consider a free scalar field theory of $\phi$. One can show 
that $e^{i \Phi (f)} \ket{\Om}$ with a smearing function $f$ supported outside of $\sO$ is equal to $e^{\Phi (\tilde f) } \ket{\Om}$ 
with $\tilde f $ complex and supported inside $\sO$; see appendix \ref{appFinal}. Note that the operator $e^{-\alpha\Phi(f)^2}$ is bounded with an (unbounded) inverse, which lies within the subset of operators we discussed in Sec.~\ref{sec3}. But $e^{-\alpha\Phi(f)}$ is not bounded, and thus lies outside. 

But the formula derived previously for the relative Tomita operator continues to hold for this case. This follows from the Bisognano-Wichmann theorem and the closure of the Tomita opertor.

As a first example, let us compute the relative modular Hamiltonian $K_{\Om f_\lam}$  in a perturbation theory in $\lam$, where $f_\lam$ corresponds to the state $e^{-\lam \Phi(f)}|\Om\ra$.
From Wick's theorem we know that $\la e^{2\lam \Phi}\ra=\exp\lb2\lam^2\la\Phi^2\ra\rb$. Therefore, 
\bea
K_{\Omega f_\lam}=-\log\lb \Delta_\Om^{1/2} e^{2\lam \Phi_J}\Delta_\Om^{1/2}\rb+2\lam^2\la \Phi^2\ra \ .
\eea
Given the exponential form of the operator $e^{-\lam_J\Phi(f)}$ and the fact that after the truncation in their spectrum the operators are bounded one can use the Baker-Campbell-Hausdorff expansion (BCH) to compute the above expression. In appendix \ref{app:calc} we compute the above logarithm using both the BCH expansion and the real-time expansion scheme introduced in Sec.~\ref{sec:perb}.
The final result is given by
\bea \label{fine}
K_{\Omega f_\lambda} &= & K_\Omega+2\lambda^2\langle\Phi(f)^2\rangle-\lambda\pi\int\frac{dt}{\text{cosh}^2(\pi t)}\Phi_J(f_t) \cr
&& \quad +\pi\lambda^2\int dt_1 dt_2\frac{g(t_2-t_1)}{\text{cosh}(\pi t_1)\text{cosh}(\pi t_2)}[\Phi_J(f_{t_1}),\Phi_J(f_{t_2})] \ .
\eea
where $\Phi(f_t)$ was defined in (\ref{ft}) and the function $g(t)$ is defined in (\ref{poi2}). Interestingly, the series expansion of $K_{\Omega f_\lam}$ terminates at order $\lam^2$ due to the fact that the commutator of fundamental free fields is proportional to the identity operator.

Similarly, one can find $K_{f_\lam g_\lam}$ where $f_\lam$ and $g_\lam$ correspond to $e^{-\lam \Phi(f)}|\Om\ra$ and $e^{-\lam \Phi(g)}|\Om\ra$, respectively. From (\ref{alphadef}) we have
\bea
\alpha_{f_\lam g_\lam}=e^{-\lam \Phi_{-i/2}(f)}e^{2\lam \Phi(g)}e^{-\lam \Phi_{i/2}(f)},
\eea
and the relative modular Hamiltonian is
\bea
K_{f_\lam g_\lam}&&=\frac{\pi\lam}{2}\int_{-\infty}^\infty \frac{dt}{\cosh^2(\pi t)}\lb \Phi(f)_{i/2}+\Phi(f)_{-i/2}-\Phi_J(g)\rb\nn\\
&&+\pi\lambda^2\int dt_1 dt_2\frac{g(t_2-t_1)}{\text{cosh}(\pi t_1)\text{cosh}(\pi t_2)}[\Phi_J(g_{t_1}),\Phi_J(g_{t_2})] \nn\\
&&+\pi\lambda^2\int dt_1 dt_2\frac{g(t_2-t_1)}{\text{cosh}(\pi t_1)\text{cosh}(\pi t_2)}[ \Phi(f_{t_1})_{i/2}+\Phi(f_{t_1})_{-i/2}, \Phi(f_{t_2})_{i/2}+\Phi(f_{t_2})_{-i/2}]\nn\ .
\eea
See Appendix~\ref{app:dfi} for a discussion of the commutator $[\Phi_{i/2}(f),\Phi_{-i/2}](g)$. 

As a second example, we consider the non-unitary squeezed state  $|\tilde{f}_2\ra=e^{-\Phi(f)^2}|\Omega\ra$. We would like to compute $K_{\Omega\tilde{f}_2}$ for this state and derive the modular flow. 
In this case, we find that the operator $\delta$ defined in (\ref{alphadef}) is
\bea
&&\de=-2\tanh\lb \lam \Phi_J(f)^2\rb \\
&& \left[\Phi_J(f)^2,\Phi_J(g)^2\right]=2\lbrace\Phi_J(f),\Phi_J(g)\rbrace[\Phi_J(f),\Phi_J(g)] \\
&&\left[\left[\Phi_J(f)^2,\Phi_J(g)^2\right],\Phi_J(h)^2\right] =\nn \\
&&4[\Phi_J(f),\Phi_J(g)]\left(\lbrace \Phi_J(f),\Phi_J(h)\rbrace[\Phi_J(g),\Phi_J(h)]+\lbrace \Phi_J(g),\Phi_J(h)\rbrace[\Phi_J(f),\Phi_J(h)]\right)\ .
\eea
Using Wick's theorem we can find the normalization of this state to be $\la e^{-2\lam \Phi^2}\ra=(1+4\lam\Phi^2)^{-1/2}$.
Then, the relative modular operator is
\bea
K_{\Om \tilde{f}_2}&=&K_\Om-\frac{1}{2}\text{log}(1+4\lambda\la\Phi^2\ra)-\lambda\pi\int\frac{dt}{\cosh^2(\pi t)}\Phi_J(f_t)^2\nn\\
&+&\frac{\pi\lam^2}{2}\int \frac{dt_0dt_1 g(t_1-t_0)}{\cosh(\pi t_0)\cosh(\pi t_1)}\lb \lbrace\Phi_J(f_{t_0}),\Phi_J(f_{t_1})\rbrace[\Phi_J(f_{t_0}),\Phi_J(f_{t_1})] \rb \nn\\
&-&{16\pi \lam^3\ov 3} \int \frac{dt_0 dt_1dt_2}{\cosh(\pi t_0)\cosh(\pi t_2)}g(t_1-t_0)g(t_2-t_1)[\Phi_J(f_{t_0}),\Phi_J(f_{t_1})]\\&&
\times \biggl( \lbrace\Phi_J(f_{t_0}),\Phi_J(f_{t_2})\rbrace[\Phi_J(f_{t_1}),\Phi_J(f_{t_2})]+\lbrace \Phi_J(f_{t_1}),\Phi_J(f_{t_2})\rbrace[\Phi_J(f_{t_0}),\Phi_J(f_{t_2})] \biggr)+O(\lam^5)\nn\ .
\eea
Similar expressions hold for $K_{\tilde{f}_2\tilde{g}_2}$.

\section{Conclusions} \label{sec:conc}

In this paper, we developed a formalism to obtain the modular and the relative modular operator of general excited states from the modular operator in the vacuum. This enables us to obtain the modular and the relative modular flow of all excited states. 
It would be interesting to apply these techniques to more examples of  excited states and more general regions of a QFT. We would also like to point out a few future directions for potential applications of our work:

\ben 

\item The study of modular flows in excited states of a conformal field theory for a half-space or spherical regions
for which the vacuum modular operators are known explicitly~\cite{Bisognano:1976za,Hislop:1981uh}.

\item  The study of modular flows for regions which are obtained from half space by small deformations in a general excited state. 
Such modular flows have played an important role in recent proofs of the average null energy condition and 
quantum null energy condition~\cite{Faulkner:2016mzt,Balakrishnan:2017bjg}. The techniques and perspectives developed here could be helpful for finding a simpler proof of the quantum null energy condition, which currently requires rather intricate replica trick discussions~\cite{Balakrishnan:2017bjg}. 

\item The study of modular flows in conformal perturbation theory (i.e. conformal field theory perturbed by a relevant operator). 

\item The study of the entanglement wedge reconstruction in holography. Furthermore, the entanglement entropy of the boundary theory for disconnected regions can undergo ``phase transitions" as one varies the size of boundary regions~\cite{Headrick:2010zt}. 
This implies that the entanglement wedge in the bulk and its modular flow patterns also undergo a similar transition.

\een

\vspace{0.2in}   \centerline{\bf{Acknowledgements}} \vspace{0.2in}
We thank Tom Faulkner, Daniel Harlow, Edward Witten and Yoh Tanimoto for discussions. This work is partially supported by the Office of High Energy Physics of U.S. Department of Energy under grant Contract Numbers DE-SC0012567 and DE-SC0019127. The work of NL is supported by a grant-in-aid from the National Science Foundation grant 
number PHY-1606531.

\appendix

\section{Tensor diagrams} \label{app:a}

In 1971 Penrose proposed a diagrammatic notation to graphically represent tensor manipulations \cite{penrose1971applications}. This tensor diagram notation was further generalized and evolved into the tensor network notation that is, nowadays, widely used in the study of multi-partite finite quantum systems and quantum information theory. See \cite{wood2011tensor} for a review. In this appendix, we use tensor network diagrams to graphically represent the statements of the Tomita-Takesaki theory. 

\begin{figure}
\centering
\includegraphics[width=0.8\textwidth]{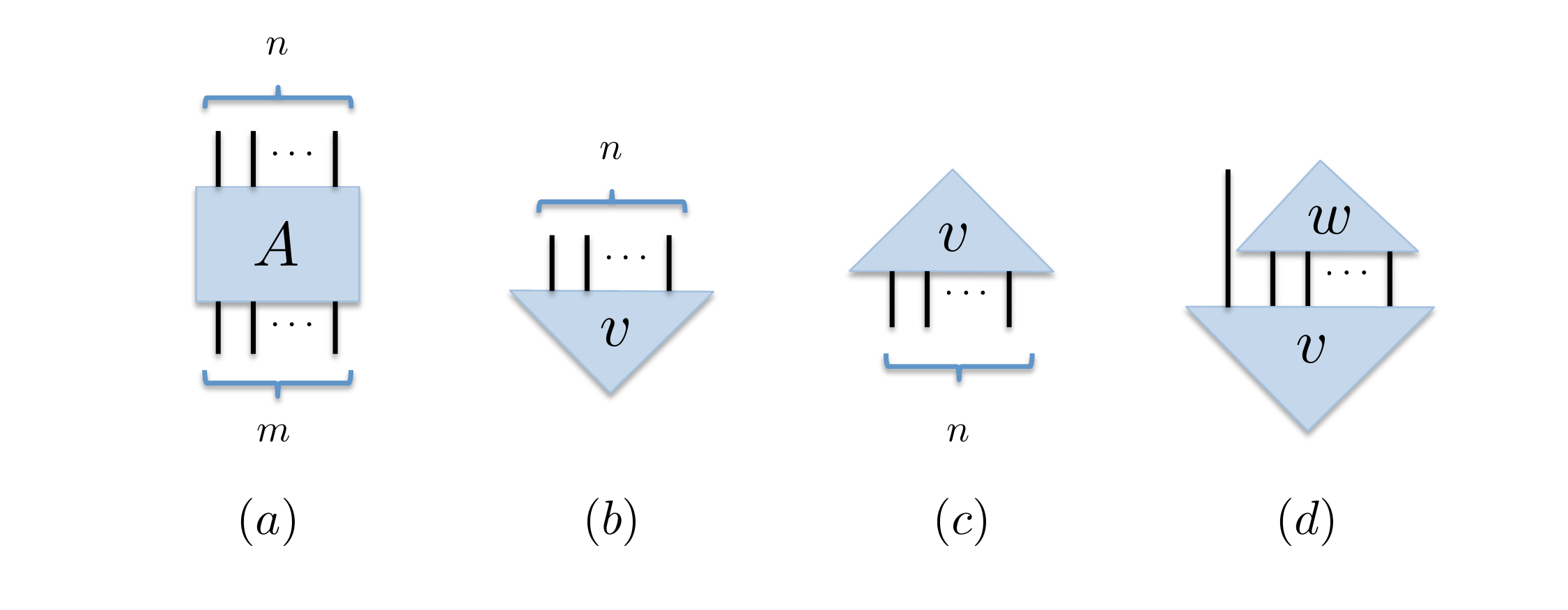}\\
\caption{\small{(a) A tensor with $m$ lines attached to the bottom and $n$ lines to the top represents a linear operator from $\cH^{\otimes m}\to \cH^{\otimes n}$ (b) A ket vector represents a quantum state $|v\ra\in\cH^{\otimes n}$ in a particular basis. (c) The dual bra $\la v|$} (d) An example of an inner product between states with different number of legs: $\la w|v\ra\in\cH$.}
\label{fig0}
\end{figure}

\begin{figure}
\centering
\includegraphics[width=0.8\textwidth]{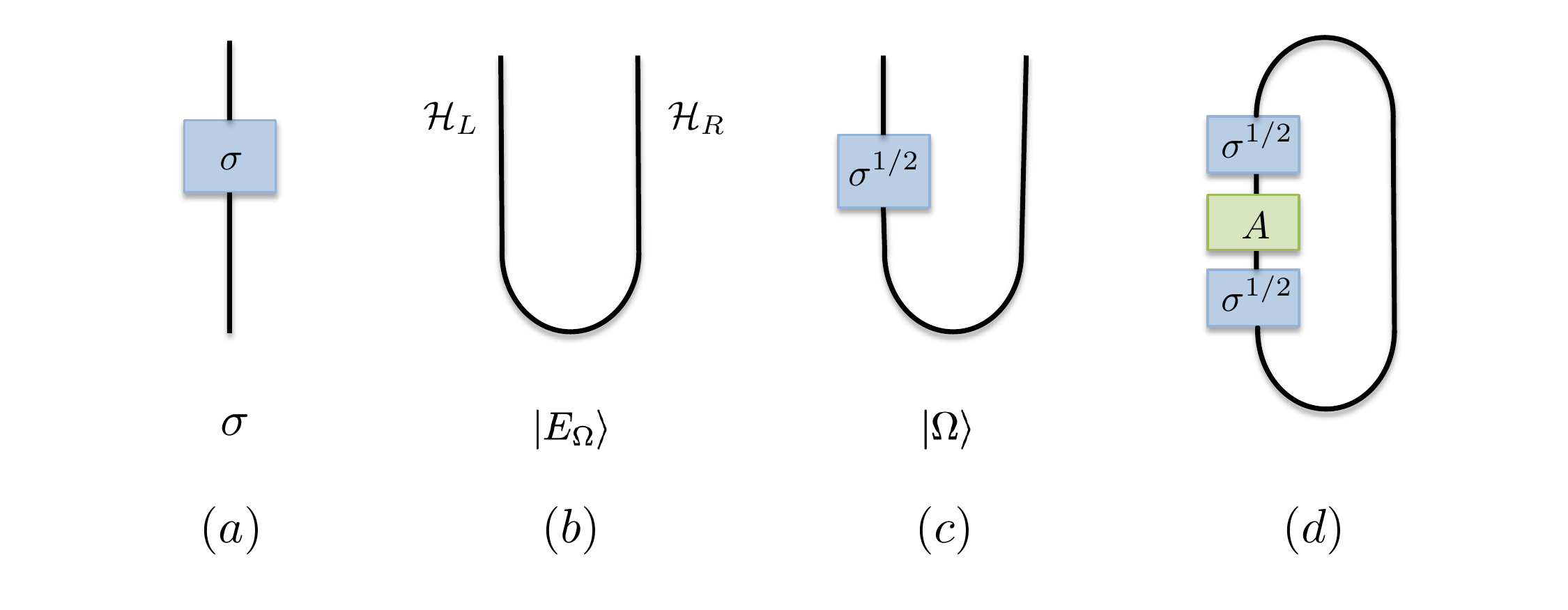}\\
\caption{\small{ Tensor diagram representation of (a) density matrix $\sigma$ (b) The unnormalized maximally entangled state built from $\sigma$ (c) The purification of $\sigma$ in a two-copy Hilbert space. (d) The expectation value of an operator in a state: $\sigma(A)=\la E_\Omega|(\sigma^{1/2}\otimes \mathcal{I})A(\sigma^{1/2}\otimes \mathcal{I})|E_\Omega\ra=tr(\sigma A)$.}}
\label{fig1}
\end{figure}

A tensor with $k$ indices can be thought of as an $k$-dimensional array of complex numbers that we depict by a box with $k$ legs attached. A box with one leg attached represents a complex vector that we choose to belong to a local Hilbert space $\cH$ of dimension $d$ written in a particular basis, see figure \ref{fig0} (b) (we often represent vectors with triangles and arrays with more indices with boxes). We pick the convention that a leg at the top of the box represents a ket vector in a Hilbert space $|\psi\ra$, and an leg to the bottom is a dual vector $\la \psi|$. Therefore, a box with a leg below and above is a linear operator $A:\cH\to \cH$. Attaching legs has the interpretation of computing the inner product of two vectors; figure \ref{fig1} (d). For simplicity, we assume all local Hilbert spaces to be isomorphic. The  implicit choice of basis picked by thinking of an array of complex numbers as linear operators in this notation is often referred to as the {\it computational basis}. A tensor with $m$ legs to the bottom and $n$ to is a linear operator  from $\cH^{\otimes m}$ to $\cH^{\otimes n}$.

The discussion below parallels the discussion in section 2.2. 
A density matrix $\sigma$ can be thought of either as a map from $\cH\to \cH$ or as its purification in a two-copy Hilbert space $\mathcal{H}_L\otimes \mathcal{H}_R$:
\bea
|\Omega\ra= \sum_{a=1}^d \sqrt{\lambda_a}|a\ra_L |a \ra_R\ .
\eea
Here, the two copies $\mathcal{H}_L$ and $\mathcal{H}_R$ are isomorphic. 
 It is convenient to use the Schmidt basis of $|\Omega\ra$ to define an unnormalized maximally entangled state
 \bea
 |E_\Omega\ra\equiv \sum_a |a\ra_L|a\ra_R
 \eea
 as we did in (\ref{EOmega}).
It is clear that $|\Omega\ra=(\sigma^{1/2}\otimes \mathcal{I}) |E_\Omega\ra$; see figure \ref{fig1}. The expectation value of an operator $A\in\mathcal{H}_L$ in density matrix $\sigma$ is $\la \Omega| A|\Omega\ra=tr(\sigma A)$. Note that the state $|E_\Omega\ra$ provides a definition of a trace for operators: $\la E_\Omega |A|E_\Omega\ra=tr(A)$. This is because $\la E_\Omega| A B|E_\Omega\ra=\la E_\Omega|B A|E_\Omega\ra$.

\begin{figure}
\centering
\includegraphics[width=0.8\textwidth]{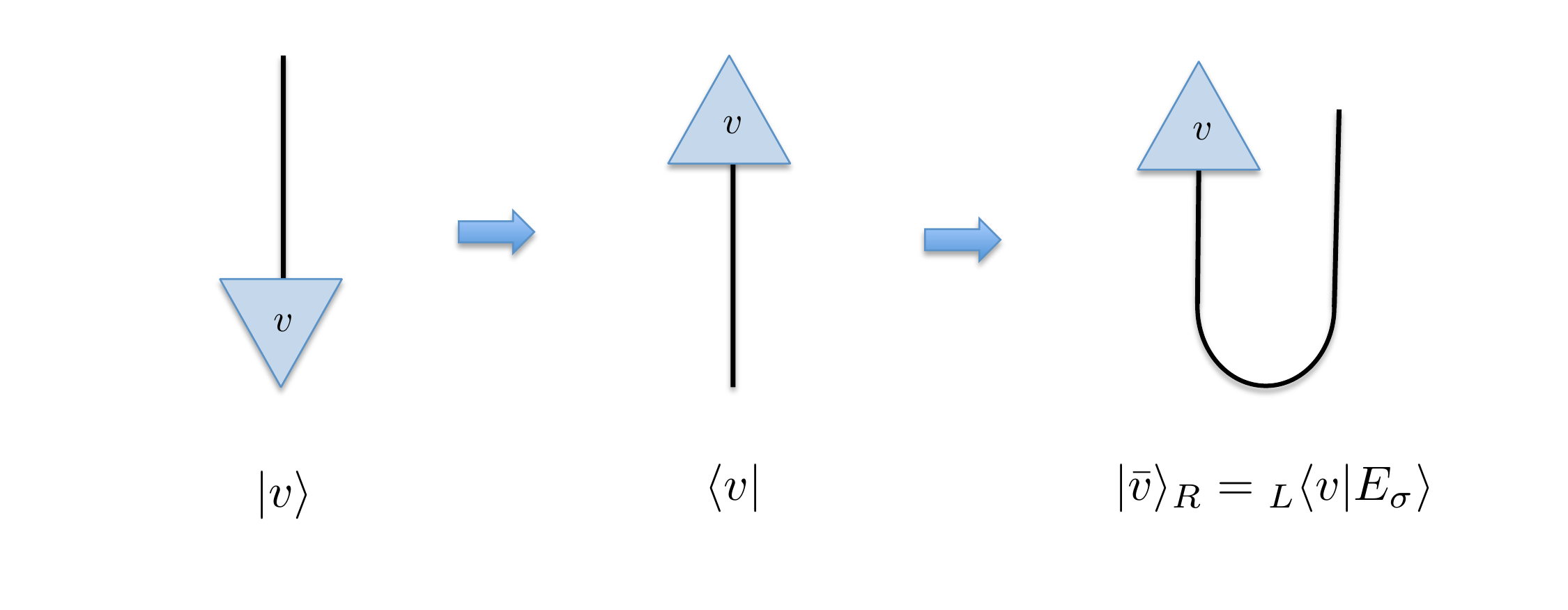}\\
\caption{\small{The definition of the anti-linear operator $T_\sigma$.}}
\label{fig2}
\end{figure}

 The purification of $\sigma$ by $|\Omega\ra$ defines for us an anti-linear map $T_\sigma:\mathcal{H}_L\to \mathcal{H}_R$:
 \bea
 &&T|v\ra_L={}_L\la v|E_\Omega\ra=|\bar{v}\ra_R\nn\\
 &&|v\ra_L=\sum_a v_a |a\ra_L\nn\\
 &&|\bar{v}\ra_R=\sum_a v_a^* |a\ra_R\ .
 \eea
 The operator $T_\sigma$ is anti-unitary because $\la T_\sigma v| T_\sigma\omega\ra=\la v|\omega\ra^*$. For anti-unitary operators the definition of adjoint is changed to $\la v|T^\dagger w\ra\equiv \la T v|w\ra^*$. Therefore, $T_\sigma=T_\sigma^\dagger=T_\sigma^{-1}$; see figure \ref{fig2}.  
 
\begin{figure}
\centering
\includegraphics[width=0.8\textwidth]{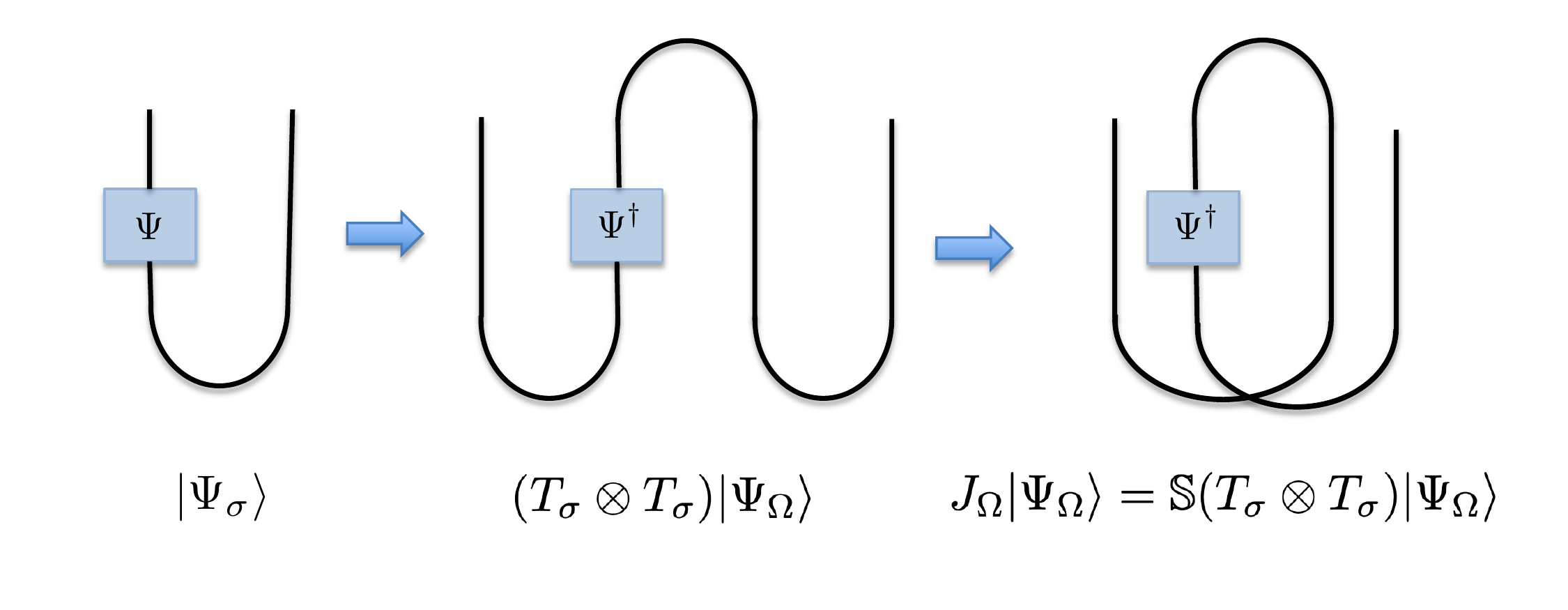}\\
\caption{\small{The definition of the anti-linear operator $J_\Omega$.}}
\label{fig3}
\end{figure}

We define the anti-linear operator $J_\Omega$ that acts on the bi-partite Hilbert space 
in the following way:
\bea
J_\Omega|v\ra_L|w\ra_R={}_R\la w|E_\Omega\ra{}_L\la v|E_\Omega\ra=|\bar{w}\ra_L|\bar{v}\ra_R
\eea
This is the same operator we introduced in (\ref{Janti}). In terms of $T_\sigma$ this is simply $J_\Omega=\mathbb{S} (T_\sigma\otimes T_\sigma)$, where the swap operator $\mathbb{S}$ is defined by $\mathbb{S}|v\ra_L|w\ra_R=|w\ra_L|w\ra_R$; see figure \ref{fig3}. Similar to $T_\sigma$, we have $J_\Omega=J_\Omega^{-1}=J_\Omega^\dagger$. It is straightforward to check the equation (\ref{JAJ0})
\bea
J_\Omega A_R J_\Omega= (T_\sigma \otimes T_\sigma)\mathbb{S} A_R \mathbb{S}(T_\sigma\otimes T_\sigma)= (T_\sigma \otimes T_\sigma)A_L(T_\sigma \otimes T_\sigma)= A^*_L
\eea
where the complex conjugate $A_L^*$ is defined with respect to the basis of $|E_\Omega\ra$. 

\begin{figure}
\centering
\includegraphics[width=0.8\textwidth]{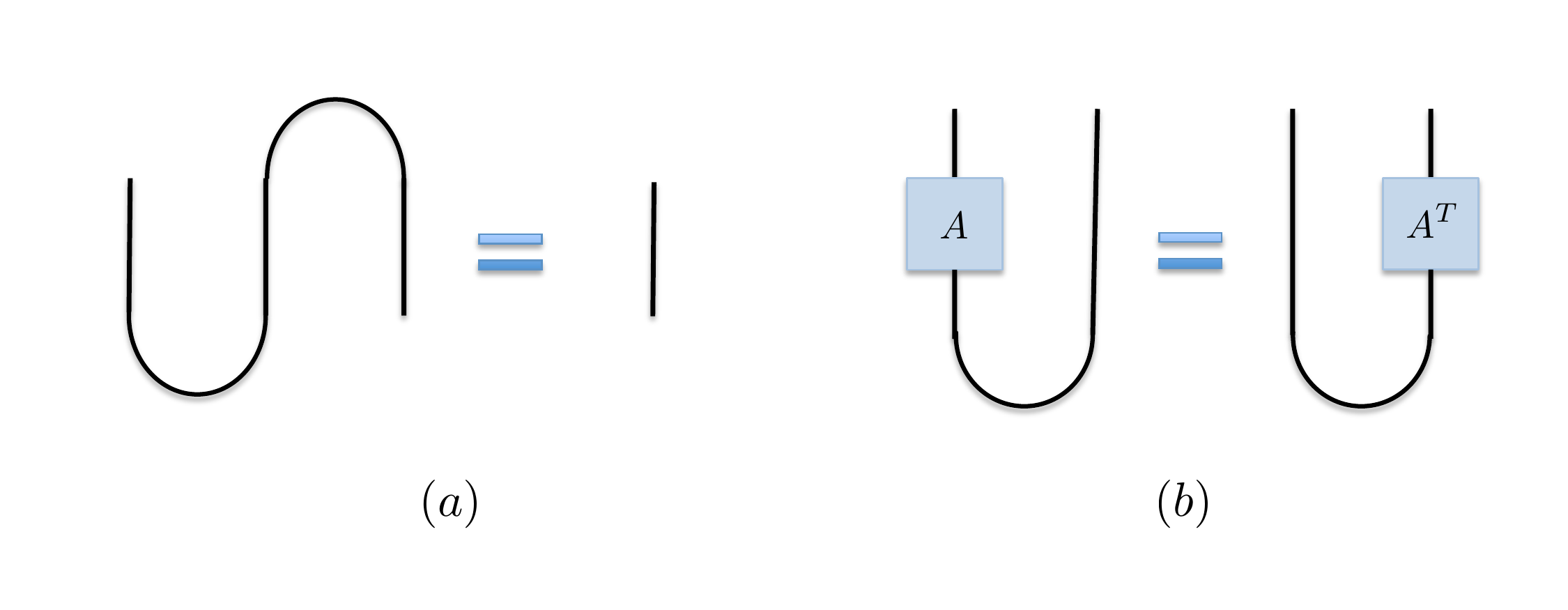}\\
\caption{\small{Tensor diagram identities.}}
\label{fig5}
\end{figure}

Figure \ref{fig5} illustrates two useful identities that we use to simplify tensor diagrams. The first identity \ref{fig5} (a) says that the inner product of the tripartite state $|E_\Omega\ra_{12}\otimes |v\ra_3$ with the bipartite state $\la E_{\Omega}|_{23}$ gives the state $|v\ra_1$:
\bea
\la E_\Omega|_{23}\lb |E_{\Omega}\ra_{12}\otimes |v\ra_3\rb= |v\ra_1\ .
\eea
If Alice holds systems $1$ and Bob holds systems $2$ and $3$, this identity says if they start sharing a maximally entangled pair, then Bob can transfer any quantum state $|v\ra$ to Alice by performing a measurement on his systems in a judiciously chosen basis. This is the idea behind the teleportation protocol.\footnote{The classical communication needed to achieve the teleportation protocol is hidden here in the basis dependence of $|E_\Omega\ra$.} The second identity \ref{fig5} (b) says that if Alice and Bob share a maximally entangled pair Bob by acting locally with an operator $A^T$ can reproduce the same effect as Alice acting with $A$. 
An advantage of the tensor network representation is that using this identities one can simplify complicated diagrams by pulling on lines to straighten them, and pushing operators through entangled pairs.

The starting point of the Tomita-Takesaki theory is the anti-linear operator $S_\Omega$ defined in (\ref{DefS}):
\bea\label{DefS2}
\forall A_L\in \mathcal{A}_L\qquad S_\Omega A_L|\Omega\ra=A_L^\dagger|\Omega\ra
\eea
where by $S_\Omega$ we mean that of the algebra $\mathcal{A}_L$.
Note that since the density matrix $\sigma$ is cyclic (full rank) the above equation defines the action of $S_\Omega$ everywhere in the Hilbert space. 

\begin{figure}
\centering
\includegraphics[width=0.8\textwidth]{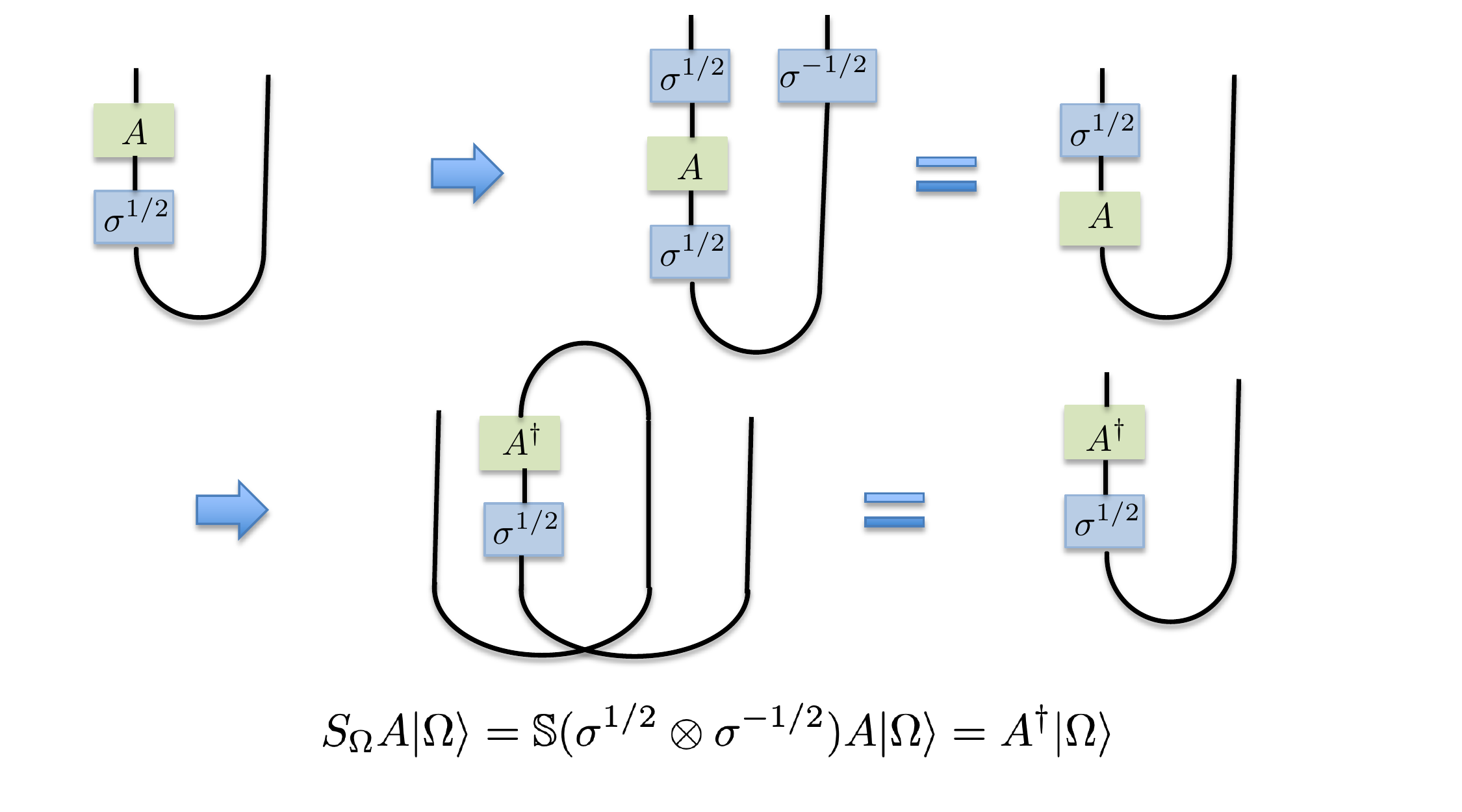}\\
\caption{\small{The action of the operator $S_\Omega$ on an arbitrary state $A|\Omega\rangle$.}}
\label{fig4}
\end{figure}

Figure \ref{fig4} uses tensor diagrams to show that $S_\Omega=J_\sigma(\sigma^{1/2}_L\otimes \sigma_R^{-1/2})$. Here, the swap operator $\mathbb{S}$ swaps two lines.
This equation can be understood as a polar decomposition of $S_\Omega$ in terms of an anti-unitary $J_\Omega$ and a positive operator $\Delta_\Omega^{1/2}$. The operator $\Delta_\Omega$ is the modular operator of $\Omega$:
\bea
\Delta_\Omega=S_\Omega^\dagger S_\Omega=\sigma_L\otimes \sigma_R^{-1}\ .
\eea
The state $\Omega$ is symmetric under the action of $S_\Omega|\Omega\ra=J_\Omega|\Omega\ra=\Delta_\Omega |\Omega\ra=|\Omega\ra$. The unitary transformations associated with this symmetry are called the modular flow, i.e. $U_\Omega(s)=\Delta_\Omega^{is}$  for real $s$.

Note that the modular operator acts on an operator in $\mathcal{H}_L$ as
\bea
\Delta_\Omega A=\sigma^{1/2}A(\sigma^{-1/2})^T
\eea
where $\sigma^T$ is the transpose of $\sigma$.

According to the GNS quantization other {\it vectors} in the Hilbert space are found by the action of $\Psi_L\in \mathcal{A}_L$ on $|\Omega\ra$:
\bea
&&|\Psi_\Omega\ra\equiv (\Psi_L\otimes \mathcal{I})|\Omega\ra\nn\\
&&\la \Psi_\Omega^\dagger|A|\Psi_\Omega\ra=tr(\psi_\sigma A)\ .
\eea
where $\psi_\sigma=\Psi\sigma\Psi^\dagger$ is another density matrix in $\mathcal{H}_L$.
Following \ref{relmoddef}, we define the relative Tomita operator $S_{\Psi\Omega}$ for any two density matrices $\psi$ and $\sigma$ by
\bea
\forall A_L\in \mathcal{A}_L\qquad S_{\Psi\Omega}A_L |\Omega\ra=A^\dagger_L |\Psi_\Omega\ra \ .
\eea
Using tensor diagrams it is straightforward to see that the operator $S_{\Psi\Omega}=J_\Omega (\Psi^\dagger_L\sigma_L^{1/2}\otimes \sigma^{-1/2}_R)$. The Hermitian operator 
\bea
\Delta_{\Psi\Omega}\equiv S_{\Psi\Omega}^\dagger S_{\Psi\sigma}= \Psi_L^\dagger\sigma_L\Psi_L\otimes \sigma_R^{-1}
\eea
is called the relative modular operator of $\Psi$ and $\Omega$.
It generates a group of unitary transformations $U_{\Psi\Omega}(s)=\Delta_{\Psi\Omega}^{is}$ as in (\ref{relm}). It is clear that when $\Psi=\Omega$ ($\Psi_L=\mathcal{I}_L$) the relative modular flow reduces to the modular flow $U_\Omega(s)$.

Finally, we compute the cocycle in (\ref{uu3}) for three density matrices $\sigma$ and
\bea
&&\psi=\Psi^\dagger \sigma\Psi\nn\\
&&\phi=\Phi^\dagger \sigma\Phi\ .
\eea
We define the unitary transformations 
\bea
&&u_{\Psi\Omega}(t)\equiv U_{\Psi\Phi}(t) U_{\Omega\Phi}^\dagger(t)= (\psi_L^{it}\otimes \phi_R^{-it})(\sigma_L^{it}\otimes \phi_R^{-it})^\dagger=(\psi_L^{i t}\sigma_L^{-is})\otimes \mathcal{I}\nn\\
&&u'_{\Psi\Omega}(t)\equiv U^\dagger_{\Phi\Psi}(t) U_{\Phi\Omega}(t)= (\phi_L^{i t}\otimes \psi_R^{-it })^\dagger(\phi_L^{is}\otimes \sigma_R^{-it})=\mathcal{I}\otimes (\psi_R^{it}\sigma_R^{-it})\ .
\eea
As opposed to modular and relative modular flows, the flow generated by the cocycle acts only on $\mathcal{H}_L$. 
 Both definitions of independent of the choice of the state $\Phi$, therefore we can choose $\Phi=\Omega$ 
 \bea\label{RN}
(D\Psi:D\Omega)(t)\equiv u_{\Psi\Omega}(t)=\Delta_{\Psi\Omega}^{i t} \Delta_\Omega^{-i t}=(\psi_L^{i t}\sigma_L^{-i t})\otimes \mathcal{I}\ .
\eea

\section{Polar decomposition}\label{AppSrivatsan}
Consider an operator $\Phi^\dagger$ and its polar decomposition $\Phi^\dagger= W|\Phi|$ with $W$ a partial isometry and $|\Phi|^2=\Phi^\dagger\Phi$. We would like to show that if $\Phi^\dagger$ is invertible $W$ is a unitary and $|\Phi|$ is also invertible. 
%
First note that an injective $\Phi^\dagger$ implies that $|\Phi|$ is injective. Since $|\Phi|$ is self adjoint and injective it  has a dense range. Therefore, $|\Phi|^{-1}$ is a densely defined linear operator:
\begin{align}
|\Phi|^{-1} = (\Phi^\dagger)^{-1}W\ .
\end{align}
If there were a non-zero vector $|x\ra\in\mathcal{H}$ such that $W|x\ra=0$ we would get $(\Phi^{\dagger})^{-1}W|x\ra=0$, which would make $|x\ra\in\text{domain}(|\Phi|^{-1})$. Since $|\Phi|^{-1}$ is injective on its domain the partial isometry $W$ is also injective, i.e., it is an isometry.
That is to say 
\begin{align}
W^{\dagger}W=\mathcal{I}_{\mathcal{H}}
\end{align}
Since $\Phi^\dagger$ has dense range then
\begin{align}
\text{range}(\Phi^\dagger) = W(\text{range}(|\Phi|))\subseteq \text{range}(W)
\end{align}
This means the range of $W$ is dense. Since the range of any isometry is always closed, this implies $W$ is surjective, hence it is a unitary.
\section{BCH for the unitary case}

A vector $x\in\mathcal{H}$ is said to be analytic for operator $T$ if the series representation for $e^{Ts}$ converges absolutely on $x$ for some $s>0$:
\begin{align}
x\in D(T^n)\hspace{20pt} \forall n>0\\
\sum_{n=0}^\infty \frac{s^n}{n!}||T^n x||<\infty,
\end{align}

Suppose $\mathcal{W}$ is a wedge. Let $\Phi(f)$ be the field operators smeared against Schwarz space functions having support inside the wedge. The common invariant domain of the $\Phi(f)$ is the set 
\begin{align}
D_0= \left\lbrace x\in\mathcal{H} : x= \Phi(f_1)...\Phi(f_n)|\Omega\rangle,\text{supp}(f_i)\in\mathcal{W}\right\rbrace.
\end{align}
This $D_0$ is dense in $\mathcal{H}$. The following is a true operator statement 
\begin{align}
U(t)\Phi(f)U^{\dagger}(t)=\Phi(f_t).\label{OpState}
\end{align}

$U(t)$ is the strongly continuous unitary generating boosts preserving the wedge. Next, Stone's theorem tells us that there is a self adjoint operator $K$ such that
\begin{align}
iK|\psi\rangle = \frac{d}{dt}U(t)|\psi\rangle_{t=0}.
\end{align}
Domain of $K$ is the set of vectors for which the RHS exists. Eq.(\ref{OpState}) implies the domain of $K$ contains at least $D_0$ and also that the following is true at least on $D_0$
\begin{align}
i[K,\Phi(f)]=\Phi(\mathcal{D}f).
\end{align}

This also implies $K D_0\subseteq D_0$, i.e., $D_0$ is invariant under $K$ (in fact, invariant under arbitrary powers of $K$ and the boost unitary $e^{it K}$). Moreover, $D_0$ is known to be analytic (in fact, entire) for each operator $\Phi(f)$ in free field theory (This is Theorem 5.2.3 in \cite{bratteli81operator}). The question now is can we series expand $e^{\lambda\Phi(f)}Ke^{-\lambda\Phi(f)}$ on $D_0$?

In general, no. Because the operator $e^{-\lambda\Phi(f)}$ typically takes us out of $D_0$ for any complex $\lambda$. For the case of unitaries in free field theory, one can argue as follows. Consider the operator
\begin{align}
e^{i\lambda\Phi(f)}e^{itK}e^{-i\lambda\Phi(f)} = e^{i\lambda\Phi(f)}e^{-i\lambda\Phi(f_t)}e^{itK}.
\end{align}
Acting this on the domain $D_0$ and using the fact that $e^{itK}D_0\subseteq D_0$, we get
\begin{align}
e^{i\lambda\Phi(f)}e^{itK}e^{-i\lambda\Phi(f)}D_0 = e^{i\lambda\Phi(f)}e^{-i\lambda\Phi(f_t)}e^{itK}D_0 = e^{i\lambda\Phi(f)}e^{-i\lambda\Phi(f_t)}D_0.
\end{align}
Next, we use the Weyl algebra to simplify this as
\begin{align}
e^{i\lambda\Phi(f)}e^{itK}e^{-i\lambda\Phi(f)}D_0 &= e^{i\lambda\Phi(f)}e^{-i\lambda\Phi(f_t)}e^{itK}D_0 = e^{i\lambda\Phi(f)}e^{-i\lambda\Phi(f_t)}D_0 \\&=\alpha(t)e^{i\lambda\Phi(f-f_t)}D_0.
\end{align}
$\alpha(t)$ is a $c$-number. Next, we prove that at least the first derivative in $t$ exists as a series expansion in $\lambda$ for the following operator
\begin{align}
e^{i\lambda\Phi(f_t)},
\end{align}
when acting on $D_0$. Here, $f$ is an arbitrary Schwarz space function with support contained in the wedge. Consider an arbitrary $x\in D_0$. Then, there is a (absolutely) convergent $\lambda$ expansion on $x$
\begin{align}
e^{i\lambda\Phi(f_t)}x = \sum_{n=0}^\infty \frac{i^n\lambda^n}{n!}\Phi(f_t)^n x.
\end{align} Next, one can Taylor expand 
\begin{align}
\Phi(f_t)^n x = \Phi(f)^n x + t \sum_{j=0}^{n-1} \Phi(f)^j \Phi(\mathcal{D}f)\Phi(f)^{n-j-1}x +R_n(t)x.
\end{align}
Here, $R_n(t)$ is a remainder term that goes to $0$ as $t\rightarrow 0$.The second term above can be rearranged to give
\begin{align}
\sum_{j=0}^{n-1} \Phi(f)^j \Phi(\mathcal{D}f)\Phi(f)^{n-j-1}x &= i[\Phi(\mathcal{D}f,\Phi(f))]\sum_{j=0}^{n-2}(n-j-1)\Phi(f)^{n-2}x+\Phi(f)^{n-1}\Phi(\mathcal{D}f)x\nn \\&= i[\Phi(\mathcal{D}f,\Phi(f))]\frac{n(n-1)}{2}\Phi(f)^{n-2}x+\Phi(f)^{n-1}\Phi(\mathcal{D}f)x.
\end{align}Thus, we get
\begin{align}
e^{i\lambda\Phi(f_t)}x &= \sum_{n=0}^\infty \frac{i^n\lambda^n}{n!}\Phi(f_t)^n x\label{Term1} \\&= \sum_{n=0}^\infty \frac{i^n\lambda^n}{n!}\Phi(f)^n x+t\sum_{n=0}^\infty\frac{i^n\lambda^n}{n!} i[\Phi(\mathcal{D}f,\Phi(f))]\frac{n(n-1)}{2}\Phi(f)^{n-2}x \label{Term2}\\&+t\sum_{n=0}^\infty\frac{i^n\lambda^n}{n!}\Phi(f)^{n-1}\Phi(\mathcal{D}f)x+\sum_{n=0}^\infty\frac{i^n\lambda^n}{n!}R_n(t)x.\label{Term3}
\end{align} In a $t$-neighbourhood of $0$, one can show that Eq.(\ref{Term1}) convergens uniformly in $t$ due to the bounds given in Theorem 5.2.3 \cite{bratteli81operator} and the Weierstrass M-test. This automatically implies all terms in Eq.(\ref{Term3}) converge unformly in $t$ which implies
\begin{align}
\text{lim}_{t\rightarrow 0}\sum_{n=0}^\infty\frac{i^n\lambda^n}{n!}R_n(t)x = 0.
\end{align} Thus, Taylor's theorem with remainder now implies the first $t$-derivative of the LHS of Eq.(\ref{Term1}) exists as a convergent power series in $\lambda$. This is sufficient to conclude
\begin{align}
e^{i\lambda\Phi(f)} K e^{-i\lambda\Phi(f)},
\end{align}
exists as a power series in $\lambda$ at least on $D_0$

\section{Analytic properties of modular evolved operators}\label{anaprop}

Consider the following operator :
\bea
I = \int_{-\infty}^\infty dt e^{-rt^2}\Phi(t),
\eea
where $\Phi\in\mathcal{A}$, $\Phi(t)=\sigma^\Omega_t(\Phi)$ is the modular flow of $\Phi$ in the state $|\Omega\rangle$.

Note that the modular flow of $I$ is simply
\begin{align*}
\sigma_{t_0}(I) = \int_{-\infty}^\infty dt e^{-r(t-t_0)^2}\Phi(t)
\end{align*} 

This makes it clear that $\sigma_{t_0}(I)$ has an entire analytic extension in $t_0$ for bounded $\Phi$, because the dominated convergence theorem justifies the differentiation with respect to $t_0$ inside the integral sign since $\Phi(t)$ is uniformly bounded in $t$.

Finally, lets discuss the case when $\Phi(t)$ is an unbounded closeable local operator. In general, the integral over a closeable operator need not be closeable. 

However, note that $\sigma_z(I)|\Omega\rangle$ is entire anlytic in $z$ by the same argument as above.

Further we note that $\sigma_z(I)$ is closeable for each $z$ : this is because the domain of its adjoint contains $\mathcal{A}_T'|\Omega\rangle$ which is dense. Thus, we conclude $\sigma_{-i/2}(I)$ is a densely defined closeable operator defined at least on $\mathcal{A}_T'|\Omega\rangle$.

\section{Unitary versus non-unitary operators in free fields}\label{appFinal}

Consider the non-unitary operator $e^{\Phi(f)}$ with $f$ supported in the right Rindler wedge. Our goal is to demonstrate that the state $e^{-\Phi(f)}|\Om\ra$ in free field theory can also be created by a unitary $e^{i\mO}|\Om\ra$ where $\mO$ is supported in a larger region than just the right wedge. 

It can be checked explicitly using the canonical commutation relations that 
\bea
e^{\Phi(f)}|\Om\ra\propto e^{i\mO}|\Om\ra,
\eea
where the operator $\mO$ is defined on a constant time slice $\Sigma:x^0=0$:
\bea
\mO=\int_\Sigma d^{d-1}x \lb \phi(x) \partial_t\gamma -\gamma(x)\partial_t\phi(x) \rb
\eea
and the $\gamma$ function is 
\bea
\gamma(x)=\int d^dy f(y) \la \lbrace\phi(x)\phi(y)\rbrace \ra
\eea
with the $y$-integral running over the whole spacetime. 
The important point is that since $\la\lbrace\phi(x)\phi(y)\rbrace\ra$ is non-zero for $y$ outside of the right wedge the support of the function $\gamma(x)$ leaks outside of the right wedge.

\section{Some calculation details} \label{app:calc}

Here, we present the calculation that leads to the expression in ~\eqref{fine} using both the BCH method and the real-time perturbation series in section \ref{sec3}. 

\subsection{BCH method}

In the family of non-unitary states discussed in section \ref{sec4}, the operators $e^{-\lam\Phi(f)}$ are exponentials of integrated local operators, hence it seems convenient to work with a perturbative $\lam$ expansion of the relative modular operator in terms of the exponents $\Phi(f)$. Such an expansion is provided by the Baker-Campbell-Hausdorff (BCH) formula. To use the BCH formula we first write the relative modular Hamiltonian as
\bea\label{useBCH}
&&K_{\Om f_\lam}=-\log\lb e^{\lam X(f)}e^Y\rb+2\lam^2\la \Phi(f)^2\ra\nn\\
&&X(f)=2(\Phi_J)_{i/2}(f), \qquad Y=-K\ .
\eea
The BCH formula written in a compact form is the following expansion
\bea\label{BCHExpan}
&&\log\lb e^A e^B\rb=A+\int_0^1 dt \:\psi\lb e^{ad_A} e^{t\: ad_B}\rb B\nn\\
&&=B+\int_0^1 dt\: \tilde{\psi}\lb e^{t\: ad_A} e^{ad_B}\rb A\nn\\
&&\psi(e^x)=e^x\tilde{\psi}(e^x)=\frac{x e^x}{e^x-1}=x+\tilde{\psi}(e^x)\nn\\
&&\tilde{\psi}(e^x)=\sum_{n=0}^\infty \frac{B_n^-}{n!} x^n\nn\\
&&\psi(e^x)=\sum_{n=0}^\infty \frac{B_n^+}{n!} x^n,\qquad B^-_n=(-1)^nB^+_n
\eea
where
\bea
ad_A B=[A,B],\qquad ad_A^n B=[A,ad_A^{n-1}B],
\eea
the coefficients $B_n^\pm$ are the Bernoulli numbers, and $\psi$ and $\tilde{\psi}$ are their generating functions.

In a free theory, the commutator of fundamental fields is 
\bea
[\Phi(f),\Phi(g)]=-i\la f,\Delta g\ra
\eea
where $\la f,\Delta g\ra$ is anti-symmetric. In particular, $\la f,\Delta f\ra=0$. From (\ref{Deltaanti-def}) for the operator $X(f)$ in (\ref{useBCH}) we find
\bea
&&ad_Y^m X(f)=(i\mathcal{D})^mX(f)\nn\ .
\eea
As a result we learn that $ad_X ad_Y^{2k}X=0$. 
Consider the BCH expansion of $\log(e^{\lam X}e^Y)$ when $ad_X ad_Y^{m}X$ is proportional to the identity operator for $m$ odd and vanishes for $m$ even. Our goal is to show that in this case the BCH expansion terminates at $O(X^2)$.

First consider the BCH expansion up to the second order in $\lam$
%
\bea
&&\log \lb e^{\lam A} e^B\rb=B+\lam\int_0^1 dt \:\sum_{n=0}^\infty\frac{B_n^-}{n!} \tilde{x}_t^n A\nn\\
&&\tilde{x}_t=\log\lb e^{t \lam\: ad_A}e^{ad_B}\rb=ad_B+t\lam\tilde{\psi}(e^{ad_{ad_B}})ad_A+O(\lam^2)\ .
\eea
Therefore,
\bea\label{bernoulli2}
\log\lb e^{\lam A} e^B\rb&=&B+\lam\tilde{\psi}(e^{ad_B})A+\frac{\lam^2}{2}\sum_{n=1}^\infty \sum_{k=0}^{n-1}\frac{B_n^-}{n!}ad_B^k\tilde{\psi}(e^{ad_{ad_B}})ad_Aad_B^{n-k-1}A+O(\lam^3)\nn\\
\eea
If $ad_Xad_Y^m X$ is a c-number as was the case for non-unitary coherent states we have 
\bea\label{commuteourcase}
&&[ad_Y,ad_X]ad_Y^m X=-ad_Xad_Y^{m+1}X\nn\\
&&[ad_Y,\cdots [ad_Y,ad_X]]ad_Y^m X=(-1)^n ad_X ad_Y^{m+n} X\nn\\
&&\tilde{\psi}(e^{ad_{ad_Y}})ad_X ad_Y^n X=\sum_{n=0}^\infty \frac{B_n^+}{n!}ad_X ad_Y^{m+n} X\ .
\eea
Hence, in the expansion (\ref{bernoulli2}) of $\log(e^{\lam X}e^Y)$ at the second order only the $k=0$ term contributes. Furthermore at order $\lam^3$ or higher in the BCH expansion, we always have at least two $ad_X$ whose action on $ad_Y^m X$ vanishes. Therefore, in this case, the BCH expansion is $\lam^2$ exact and given by
\bea\label{BCHsecondary}
&&\log(e^{\lam X}e^Y)=Y+\lam\tilde{\psi}(e^{ad_Y})X+\frac{\lam^2}{2}ad_X\mathcal{F}(ad_Y)X\nn\\
&&\mathcal{F}(x)=\sum_{n=0}^\infty \sum_{m=1}^{\infty}\frac{B_m^-B_n^+}{n!m!}x^{m+n-1}=\frac{e^x}{e^x-1}\lb \frac{x}{e^x-1}-1\rb=\frac{x+1-e^x}{4\sinh^2(x/2)}\ .
\eea  
As we discussed earlier, we can drop the terms in $\mathcal{F}(x)$ that have even powers of $x$. 

Since the commutator of $[\Phi(f),(i\mathcal{D})^m\Phi(f)]$ is proportional to the identity operator the BCH expansion for the relative modular operator truncates at order $\lam^2$ and becomes
\bea\label{BCHsecond}
&&K_{\Om f_\lam}-K_\Om=-2\lam\tilde{\psi}(e^{ad_{-K}})(\Phi_J)_{i/2}(f)-2\lam^2\left[(\Phi_J)_{i/2}(f),\mathcal{F}(ad_{-K})(\Phi_J)_{i/2}(f)\right]+2\lam^2\la \Phi(f)^2\ra\nn\\
\eea  
Let us start with the term linear in $\lam$. We use the spectral decomposition of $\Delta$: 
\bea\label{spec}
\Delta=\int_{-\infty}^\infty e^{-\omega}\: P(d\omega),
\eea
with $P(d\omega)$ a projection-operator-valued-measure (POVM). We write the first order term in $\lam$ as
\bea
\tilde{\psi}(e^{ad_{-K}})X_{i/2}&=&\int_{-\infty}^\infty d\omega d\omega' \frac{(\omega-\omega')}{1-e^{-(\omega-\omega'})}P(d\omega)X_{i/2} P(d\omega')\nn\\
&=&\int_{-\infty}^\infty d\omega d\omega' \frac{(\omega-\omega')}{2\sinh\lb\frac{\pi \omega}{2}\rb}P(d\omega)X P(d\omega')\nn\\
&=&\int_{-\infty}^\infty  d\omega d\omega' \int_{-\infty}^\infty dt \: \frac{\pi e^{i(\omega-\omega')t}}{2\cosh^2(\pi t)} P(d\omega)XP(d\omega')\nn\\
&=&\frac{\pi}{2}\int\frac{dt}{\cosh^2(\pi t)} X(t)\ .
\eea 
Plugging $X=-2\Phi_J$ in the above expression we find that the first order correction to the relative modular operator is
\bea\label{firstK}
K^{(1)}_{\Om f_\lam}=-\pi \int \frac{dt}{\cosh^2(\pi t)}\Phi_J(f_t)\ .
\eea

Now, consider the second order term:
\bea\label{diagonalmatrix}
\left[(\Phi_J)_{i/2}(f),\mathcal{F}(ad_{-K})(\Phi_J)_{i/2}(f)\right]&=&\int d\omega d\omega' d\omega''\: P(d\omega)\Phi_J P(d\omega')\Phi_J P(d\omega'') \nn\\
&&\times e^{-(\omega-\omega'')/2} \lb \mathcal{F}(\omega''-\omega')-\mathcal{F}(\omega'-\omega)\rb
\eea
Since we are only considering free fields we focus on the part of the above commutator that is proportional to the identity operator by setting $\omega=\omega''$. This ``diagonal term" is
\bea
\int d\omega d\omega' \:\frac{(\omega-\omega')-\sinh(\omega-\omega')}{1-\cosh(\omega-\omega')}\: P(d\omega)\Phi(f)P(d\omega')\Phi(f)P(d\omega)\ .
\eea
We use the identity
\bea
\frac{(\omega-\omega')-\sinh(\omega-\omega')}{1-\cosh(\omega-\omega')}=\frac{-\pi }{2}\int_{-\infty}^\infty \frac{dt ds}{\cosh(\pi t)\cosh(\pi s)}g(s-t) \lb e^{i(\omega-\omega')(s-t)}-e^{-i(\omega-\omega')(s-t)}\rb\nn
\eea
to write the (\ref{diagonalmatrix}) as
\bea\label{secondK}
-\frac{\pi}{2}\int_{-\infty}^\infty \frac{dt ds}{\cosh(\pi t)\cosh(\pi s)}g(s-t) [\Phi_J(t),\Phi_J(s)]\ .
\eea
Plugging (\ref{firstK}) and (\ref{secondK}) back into (\ref{BCHsecond}) we obtain
\bea
K_{\Om f_\lam}-K_\Om&=&2\lam^2\la \Phi(f)^2\ra-\pi \lam \int \frac{dt}{\cosh^2(\pi t)}\Phi_J(f_t)\nn\\
&+& \pi\lam^2\int_{-\infty}^\infty \frac{dt ds}{\cosh(\pi t)\cosh(\pi s)}g(s-t) [\Phi_J(t),\Phi_J(s)]
\eea
as was promised in section \ref{sec4}.

\subsection{Real-time method}\label{RT}
Now, we reproduce the BCH answer above using the perturbation theory in $\lam$ discussed in section \ref{sec3}. We show that there are non-trivial cancellations between the commutator terms and contact terms that make all terms of order $\lam^3$ and higher vanish. 

From the analysis of section \ref{sec3}, for states $|f_\lam\ra= e^{-\lam\Phi(f)}$, we have
 \bega
S_{\Omega f_\lam}=e^{\lam \Phi(f)}S_\Om, \qquad
 \Delta_{\Omega f_\lam}=\Delta_\Omega^{1/2}e^{2\lambda\Phi_J}\Delta^{1/2}, \qquad 
 \alpha_{\Omega f_\lam}=1-e^{2\lam \Phi(f)_J},  \\
\delta=2\frac{1-e^{2\lambda\Phi_J}}{1+e^{2\lambda\Phi_J}} =-2\tanh\lb\lam\Phi_J(f)\rb 
=-2\lam\Phi_J(f)+\frac{2\lambda^3}{3} \Phi_J(f)^3-\frac{4\lambda^5}{15} \Phi_J(f)^5+\cdots  \ .
\end{gather}
Using the result of Sec.~\ref{sec:perb} we expand the relative modular operator as 
\bea
K_{\Om f_\lam}&=&K_\Om+2\lam^2\la\Phi(f)^2\ra+\sum_{m=1}^\infty Q_m\ .
\eea
We need the following commutators:
\bega
[\delta(t_1),\delta(t_2)]=4\lambda^2[\Phi_J(f_{t_1}),\Phi_J(f_{t_2})]-\frac{4\lam^4}{3}\lb [\Phi_J^3(f_{t_0}),\Phi_J(f_{t_1})]+ [\Phi_J(f_{t_0}),\Phi_J^3(f_{t_1})]\rb +O(\lam^6), \nn\\
[[\delta(t_1),\delta(t_2)],\delta(t_3)]=\frac{8\lam^5}{3}\lb [[\Phi_J(f_{t_1}),\Phi_J^3(f_{t_2})],\Phi_J(f_{t_3})]+ [\Phi_J(f_{t_1})^3,\Phi_J(f_{t_2})],\Phi_J(f_{t_3})]]\rb+O(\lam^7) \nn\ .
\end{gather}
Other nested commutators that appear in the expansion can be checked to be $O(\lambda^6)$ and higher. 
To find the contact terms we also need the following terms:
\bea
&&\delta(t)^3 = -8\lambda^3\Phi_J(f_t)^3+8\lambda^5\Phi_J(f_t)^5+O(\lambda^6), \\
&&\left\lbrace\delta(t_1),[\delta(t_1),\delta(t_2)^2]\right\rbrace=16\lambda^4\left\lbrace\Phi_J(f_{t_1})[\Phi_J(f_{t_1}),\Phi_J(f_{t_2})^2]\right\rbrace \ .
\eea

Then, the first term in the expansion for the relative modular Hamiltonian is the operator
\begin{align}
Q_1 &= \frac{\pi}{2}\int\frac{dt}{\text{cosh}^2(\pi t)}\delta(t)\nn\\
&=\frac{\pi}{2}\int\frac{dt}{\text{cosh}^2(\pi t)}\biggl(-2\lambda\Phi_J(f_t)+\frac{2\lambda^3}{3}\Phi_J(f_t)^3-\frac{4\lambda^5}{15}\Phi_J(f_t)^5+O(\lam^6)\biggr)\ .
\end{align}
The next term is
\begin{align}
Q_2 &=\frac{\pi}{4}\int dt_1 dt_2\frac{g(t_2-t_1)}{\text{cosh}(\pi t_1)\text{cosh}(\pi t_2)}\biggl(4\lambda^2[\Phi_J(f_{t_1}),\Phi_J(f_{t_2})]-\frac{8\lambda^4}{3}[\Phi_J(f_{t_1}),\Phi_J(f_{t_2})^3]+O(\lam^6)\biggr)\nn
\end{align}
Putting these together we already reproduce the result 
\bea
K_{\Om f_\lam}-K_\Om&=&2\lam^2\la \Phi(f)^2\ra-\pi \lam \int \frac{dt}{\cosh^2(\pi t)}\Phi_J(f_t)+\nn\\
&+& \pi\lam^2\int_{-\infty}^\infty \frac{dt ds}{\cosh(\pi t)\cosh(\pi s)}g(s-t) [\Phi_J(t),\Phi_J(s)]+O(\lam^3)\ .
\eea

Now, we need to show that all higher order terms in $\lam$ cancel for free fields.
In this case, the commutator of fundamental fields is proportional to the identity operator, hence we obtain
\begin{align}
Q_2^{free} &=\int dt_1 dt_2\frac{g(t_2-t_1)}{\text{cosh}(\pi t_1)\text{cosh}(\pi t_2)}\biggl(\pi\lambda^2[\Phi_J(f_{t_1}),\Phi_J(f_{t_2})]-2\pi\lambda^4[\Phi_J(f_{t_1}),\Phi_J(f_{t_2})]\Phi_J(f_{t_2})^2+O(\lam^6)\biggr)\nn\ .
\end{align}

The terms $Q_m$ with $m>2$ in (\ref{kiu}), in addition to the nested commutators $I_m$ also include contact terms that we denoted by $P_m$. The third term $Q_3$ splits according to
\begin{align}
Q_3&=I_3+P_3\nn\\
I_3&= \frac{4\pi\lambda^5}{9}\int dt_1 dt_2 dt_3\frac{g(t_2-t_1)g(t_3-t_2)}{\text{cosh}(\pi t_1)\text{cosh}(\pi t_3)}\biggl([[\Phi_J(f_{t_1}),\Phi_J(f_{t_2})^3],\Phi_J(f_{t_3})]+[[\Phi_J(f_{t_1})^3,\Phi_J(f_{t_2})],\Phi_J(f_{t_3})]\biggr)\nn\\
&+O(\lam^6)\nn\\
P_3&=\frac{\pi}{24}\int\frac{dt}{\text{cosh}^2(\pi t)}\biggl[-8\lambda^3\Phi_J(f_t)^3+8\lambda^5\Phi_J(f_t)^5+O(\lam^6)\biggr]
\end{align}

Specialising to free field theory, we obtain the following simplification for $Q_3$:
\begin{align}
I_3^{free}&=\frac{8\pi\lambda^5}{3}\int dt_1 dt_2 dt_3\frac{g(t_2-t_1)g(t_3-t_2)}{\text{cosh}(\pi t_1)\text{cosh}(\pi t_3)}\biggl(\Phi_J(f_{t_2})[\Phi_J(f_{t_1}),\Phi_J(f_{t_2})][\Phi_J(f_{t_2}),\Phi_J(f_{t_3})]\nn\\
&+\Phi_J(f_{t_1})[\Phi_J(f_{t_1}),\Phi_J(f_{t_2})][\Phi_J(f_{t_1}),\Phi_J(f_{t_3})]\biggr)\nn\\
P_3^{free}&=\frac{\pi}{3}\int\frac{dt}{\text{cosh}^2(\pi t)}\biggl[-\lambda^3\Phi_J(f_t)^3+\lambda^5\Phi_J(f_t)^5\biggr]
\end{align}

The same analysis can be applied to $Q_4$:
\begin{align}
Q_4&=I_4+P_4\\
I_4&=O(\lambda^6)\nn\\
P_4&=\frac{\pi}{2}\lambda^4\int dt_1 dt_2\frac{g(t_2-t_1)}{\text{cosh}(\pi t_1)\text{cosh}(\pi t_2)}\biggl(\Phi_J(f_{t_2})[\Phi_J(f_{t_1}),\Phi_J(f_{t_2})^2]+[\Phi_J(f_{t_1}),\Phi_J(f_{t_2})^2]\Phi_J(f_{t_2})\biggr)\nn\\
&+O(\lam^6)
\end{align}
which also simplify for free fields
\begin{align}
P_{4}^{free}&=2\pi\lambda^4\int dt_1 dt_2\frac{g(t_2-t_1)}{\text{cosh}(\pi t_1)\text{cosh}(\pi t_2)}\biggl(\Phi_J(f_{t_2})^2[\Phi_J(f_{t_1}),\Phi_J(f_{t_2})]\biggr)\ .
\end{align}

Finally, the last term that can contribute to order $\lam^5$ is
\begin{align}
Q_5&=I_5+P_5\nn\\
I_5&=O(\lambda^7)\nn\\
P_5&=\frac{\pi}{40}\int \frac{g(t_2-t_1)g(t_3-t_2)}{\text{cosh}(\pi t_1)\text{cosh}(\pi t_3)}\biggl[\delta(t_1)^3\delta(t_2)\delta(t_3)+\delta(t_2)\delta(t_1)\delta(t_2)^2\delta(t_3)+\delta(t_1)\delta(t_3)\delta(t_2)\delta(t_1)^2\nn\\
&+\delta(t_1)\delta(t_2)\delta(t_3)^3+\delta(t_1)\delta(t_2)^2\delta(t_3)\delta(t_2)+\delta(t_1)^2\delta(t_2)\delta(t_3)\delta(t_1)-\delta(t_1)\delta(t_2)\delta(t_3)\delta(t_1)^2\nn\\
&-\delta(t_2)^2\delta(t_3)\delta(t_2)\delta(t_1)-\delta(t_1)^2\delta(t_3)\delta(t_2)\delta(t_1)-\delta(t_1)\delta(t_2)\delta(t_3)\delta(t_2)^2+\delta(t_1)\delta(t_2)^3\delta(t_3)\biggr]\nn
\end{align}

Specializing to free field theory, we find
\begin{align}
P_5^{free}&=\frac{-8\pi\lambda^5}{3}\int dt_1 dt_2 dt_3\frac{g(t_2-t_1)g(t_3-t_2)}{\text{cosh}(\pi t_1)\text{cosh}(\pi t_3)}\Phi_J(f_{t_1})[\Phi_J(f_{t_1}),\Phi_J(f_{t_2})][\Phi_J(f_{t_1}),\Phi_J(f_{t_3})]\nn\\
&+\frac{8\pi\lambda^5}{3}\int \frac{g(t_2-t_1)g(t_3-t_2)}{\text{cosh}(\pi t_1)\text{cosh}(\pi t_3)}\Phi_J(f_{t_2})[\Phi_J(f_{t_2}),\Phi_J(f_{t_3})][\Phi_J(f_{t_2}),\Phi_J(f_{t_1})]\nn\\
&-\frac{\pi\lambda^5}{5}\int dt\frac{\Phi_J(f_t)^5}{\text{cosh}^2(\pi t)}+O(\lam^6)\nn\ .
\end{align}
Putting all this together we find that all the $O(\lambda^3,\lambda^4,\lambda^5)$ terms cancel explicitly.

\section{Commutator of Euclidean evolved operators} \label{app:dfi}

In the expression for the relative modular Hamiltonian $K_{\Phi\Om}$ we found in section \ref{sec3} various commutators of $\phi^{(1)}_{\frac{i}{2}}$ and $\phi^{(1)}_{-\frac{i}{2}}$ appear; see (\ref{p1}). It is clear that
\bea
\left[\phi^{(1)}_{\pm\frac{i}{2}}(t),\phi^{(1)}_{\pm \frac{i}{2}}(s)\right]=\Delta^{\pm\frac{i}{2}}\left[\phi^{(1)}(t),\phi^{(1)}(s)\right]\Delta^{\mp\frac{i}{2}}\ .
\eea
However, the commutators $\left[\phi^{(1)}_{\pm\frac{i}{2}}(t),\phi^{(1)}_{\mp \frac{i}{2}}(s)\right]$ are more subtle.
In this appendix, we comment on this type of commutators using the spectral decomposition of the modular operator. 

The Euclidean commutator 
\bea\label{euclid}
[\Phi(f)_{i/2},\Phi(g)_{-i/2}]=\int d\omega d\omega' d\omega'' \lb e^{\omega'-\frac{\omega}{2}-\frac{\omega''}{2}}-e^{-\lb\omega'-\frac{\omega}{2}-\frac{\omega''}{2}\rb} \rb P(d\omega)\Phi(f)P(d\omega')\Phi(g)P(d\omega'')\nn
\eea
is convergent for a nice enough function $f$. 
Naively, one might have thought that the commutator (\ref{euclid}) can be computed by comparing the perturbative expansion of section \ref{sec3} for unitary states $e^{i\lam \Phi(f)}|\Om\ra$ with the exact answer from the BCH expansion.
%
However, as we show below, formal manipulation of the commutator, indeed, reproduces the same result for the relative modular Hamiltonian of unitary states we obtained in section \ref{sec3}:
\bea
K_{e^{i\lam \Phi(f)}|\Omega}=e^{i \lam \Phi(f)}K e^{-i\lam \Phi(f)}=K+i \lam [\log\Delta,\Phi(f)]-\frac{(i\lam)^2}{2}[[\log\Delta,\Phi(f)],\Phi(f)]]+O(\lam^3),
\eea
In free theories, from the fact that commutators of fundamental fields is central we expect the commutator $[\Phi(f)_{i/2},\Phi(g)_{-i/2}]$ to be proportional to the identity operator:
\bea
-2\int d\omega d\omega' \sinh(\omega-\omega') P(d\omega)\Phi(f)P(d\omega')\Phi(g)P(d\omega)
\eea
This is simply the Fourier transform of the statement that we are considering commutators of Euclidean evolved operators and leads to no insight about them.

We again stress these are formal manipulations and we are not concerned with domain questions and closeablity in the following. Therefore, in this appendix, we do not perform the smearing we discussed earlier in Appendix \ref{anaprop}.

To obtain intuition about how this commutator reproduces the correct answer for unitary states we go through this example step by step.
Consider a unitary state $U=e^{i\lam \Phi(f)}$. The relative modular operator 
\bea
\Delta_{U|\Omega}=e^{i\lam \Phi(f)}\Delta e^{-i\lam \Phi(f)}
\eea
Define
\bea
&&a=\lb e^{i\lam \Phi(f)}\rb_{-i/2} \lb e^{-i\lam \Phi(f)}\rb_{i/2} \nn\\
&&\Phi_{i/2}\equiv \Delta^{1/2} \Phi \Delta^{-1/2}\ .
\eea
Since the state $U$ creates is already normalized we expand $a$ in $\lam$ to find
\bea
&&a=1+\lam a_1+\frac{\lam^2}{2}a_2+O
(\lam^3)\nn\\
&&a_1=i \lb \Phi_{-i/2}(f)-\Phi_{i/2}(f)\rb\nn\\
&&a_2=- \Phi^2_{-i/2}(f)-\Phi^2_{i/2}(f)+2\Phi_{-i/2}(f)\Phi_{i/2}(f)
\eea
Following the notation in section \ref{sec3} we expand $\delta$ in $\lam$
\bea
&&\delta=\frac{2(1-a)}{1+a}=-\lam a_1+\frac{\lam^2}{2}(a_1^2-a_2)+O(\lam^3)\nn\\
&&=\lam \delta^{(1)}+\frac{\lam^2}{2}\delta^{(2)}+O(\lam^3)\nn\\
&&\delta^{(1)}=i\lb \Phi_{i/2}(f)-\Phi_{-i/2}(f)\rb\nn\\
&&\delta^{(2)}=[\Phi_{i/2}(f),\Phi_{-i/2}(f)]\ .
\eea
Since $\delta$ starts at order $\lam$ then if we are interested in the modular Hamiltonian up to the second order in $\lam$ we only need
\bea
&&Q_0=\frac{\pi}{2}\int\frac{dt}{\cosh^2(\pi t)}\delta(t)=\int_0^\infty d\beta \frac{\Delta^{1/2}}{\Delta+\beta}\delta\frac{\Delta^{1/2}}{\Delta+\beta}\nn\\
&&Q_1=\frac{\pi}{4}\int \frac{dt ds}{\cosh(\pi t)\cosh(\pi s)}g(s-t)[\delta(t),\delta(s)]\nn\\
&&=\int_0^\infty d\beta \frac{\Delta^{1/2}}{\Delta+\beta}\delta\frac{\Delta-\beta}{2(\Delta+\beta)}\delta\frac{\Delta^{1/2}}{\Delta+\beta}
\eea
where we have used the following two integrals
\bea
&&\frac{\Delta^{\ha}}{\Delta+\beta}=\int_{-\infty}^\infty \frac{\beta^{it-\ha}}{2\cosh(\pi t)}\Delta^{-i t}\nn\\
&&\frac{\Delta-\beta}{2(\Delta+\beta)}=\int_{-\infty}^\infty dt\beta^{i t} g(t)\Delta^{-it}\ .
\eea
Expanding the relative modular Hamiltonian we have
\bea
&&K_{U\Omega}=K_\Omega+\lam K^{(1)}+\frac{\lam^2}{2}K^{(2)}+O(\lam^3)\nn\\
&&K^{(1)}=\frac{\pi}{2}\int \frac{dt}{\cosh^2(\pi t)}\delta^{(1)}(t)\nn\\
&&K^{(2)}=\frac{\pi}{2}\int \frac{dt}{\cosh^2(\pi t)}\delta^{(2)}(t)+\frac{\pi}{2}\int \frac{dt ds}{\cosh(\pi t)\cosh(\pi s)}g(s-t)[\delta^{(1)}(t),\delta^{(1)}(s)]\nn
\eea
We would like to compare this answer with the result we found in (\ref{nested2})
\bea\label{nestedapp}
 &&K_{U\Omega}=e^{i\lam \Phi}K e^{-i\lam \Phi}=K+i \lam [\log\Delta,\Phi(f)]-\frac{(i\lam)^2}{2}[[\log\Delta,\Phi(f)],\Phi(f)]]\nn\\
 &&=K-\lam\Phi(\mathcal{D}f)-\frac{\lam^2}{2}\la f,\Delta \mathcal{D}f\ra
\eea
where $\mathcal{D}=x^1\p_0-x^0\p_1$ is the generator of boost.

 The operator $K^{(1)}$ written using the spectral decomposition of $\Delta$ in (\ref{spec}) is
\bea
&&K^{(1)}=\frac{i\pi}{2}\int \frac{dt}{\cosh^2(\pi t)} \lb e^{i K(t+i/2)} \Phi(f) e^{-i K(t+i/2)}-e^{i K(t-i/2)} \Phi(f) e^{-i K(t-i/2)}\rb \nn\\
&&=\frac{i \pi}{2}\int d\omega d\omega'\int\frac{dt}{\cosh^2(\pi t)} \lb e^{i(t+i/2)(\omega-\omega')}-e^{i(t-i/2)(\omega-\omega')} \rb P(d\omega)\Phi(f) P(d\omega') \nn\\
&&=-i\int d\omega d\omega'(\omega-\omega') P(d\omega)\Phi(f) P(d\omega')=i[\Phi(f),K]
\eea
where we have used the fact $\int d\omega P(\omega)=\mathcal{I}$. This is the correct answer at order $\lam$. At the second order we expect to find
\bea
-[[K,\Phi(f)],\Phi(f)]]=\int d\omega d\omega' d\omega'' (2\omega'-\omega-\omega'') P(d\omega)\Phi(f)P(d\omega')\Phi(f)P(d\omega'')\ .
\eea
We show this explicitly below.
%

At the second order in $\lam$ we have two terms, first
\bea
&&\frac{\pi}{2}\int \frac{dt}{\cosh^2(\pi t)}\delta^{(2)}=\frac{\pi}{2}\int d\omega d\omega' d\omega''\int \frac{dt}{\cosh^2(\pi t)}\:P(d\omega)\Phi(f)P(d\omega')\Phi(f)P(d\omega'')\nn\\
&&\times \lb  e^{\omega'} e^{i \omega(t+i/2)} e^{-i \omega''(t-i/2)}-e^{-\omega'} e^{i \omega(t-i/2)} e^{-i \omega''(t+i/2)} \rb\nn\\
&&=\int d\omega d\omega' d\omega'\:P(d\omega)\Phi(f)P(d\omega')\Phi(f)P(d\omega'') (\omega-\omega'')\lb \frac{e^{\omega'}}{e^\omega-e^{\omega''}}+\frac{e^{-\omega'}}{e^{-\omega}-e^{-\omega''}}\rb\nn
\eea
and the second term 
\bea\label{secondterm2}
&&\int_0^\infty d\beta \frac{\Delta^{1/2}}{\Delta+\beta}\delta^{(1)}\frac{\Delta-\beta}{\Delta+\beta}\delta^{(1)}\frac{\Delta^{1/2}}{\Delta+\beta}\nn\\
&&=-\int_0^\infty d\beta\int d\omega d\omega' d\omega'' P(d\omega)\Phi(f)P(d\omega')\Phi(f)P(d\omega'') \lb \frac{e^{-\omega/2}}{(e^{-\omega}+\beta)}\frac{(e^{-\omega'}-\beta)}{(e^{-\omega'}+\beta)}\frac{e^{-\omega''/2}}{(e^{-\omega''}+\beta)}\rb\nn\\
&&\times \lb e^{-\omega/2+\omega''/2}-e^{\omega/2+\omega''/2-\omega'}-e^{-(\omega/2+\omega''/2-\omega')}+e^{\omega/2-\omega''/2}\rb\ .
\eea
The $\beta$ integral can be performed explicitly to give
\bea
&&\int_0^\infty d\beta \lb \frac{e^{-\omega/2}}{(e^{-\omega}+\beta)}\frac{(e^{-\omega'}-\beta)}{(e^{-\omega'}+\beta)}\frac{e^{-\omega''/2}}{(e^{-\omega''}+\beta)}\rb=\frac{1}{4}\csch\lb \frac{\omega-\omega'}{2}\rb\csch\lb \frac{\omega'-\omega''}{2}\rb  (\omega+\omega''-2\omega')\nn\\
&&+(\omega-\omega'')\frac{\sinh\lb \frac{\omega+\omega''-2\omega'}{2}\rb}{\sinh(\omega-\omega')-\sinh(\omega-\omega'')+\sinh(\omega'-\omega'')}\ .
\eea
Plugging this back in (\ref{secondterm2}) and adding both $\lam^2$ terms we find
\bea
&&\int d\omega d\omega' d\omega'\:P(d\omega)\Phi(f)P(d\omega')\Phi(f)P(d\omega'') (2\omega'-\omega-\omega'')=2\Phi(f)K\Phi(f)-K\Phi(f)^2-\Phi(f)^2K\nn\\
&&=-[[K,\Phi(f)],\Phi(f)]]
\eea
which is consistent with the result in (\ref{nestedapp}).


\section{Domain issues}
Consider the relative modular operator between $\mathcal{O}|\Omega\rangle$ and $|\Omega\rangle$.
\subsection{Bounded $\mathcal{O}$}
First consider $\mathcal{O}\in\mathcal{A}$. Then, we have with $S_0 = \bar{S}_\Omega \mathcal{O}^*$, $S_0^\dagger = \mathcal{O}S_\Omega^*$, $a\in\mathcal{A}$ and $a'\in\mathcal{A}'$,
\begin{align*}
S_0[a|\Omega\rangle] &= a^*\mathcal{O}|\Omega\rangle\\
S_0^\dagger[a'|\Omega\rangle] &= a^{'*}\mathcal{O}|\Omega\rangle
\end{align*} as needed. Thus, we conclude $S_0\subset \bar{S}_{\mathcal{O}|\Omega}$ and $S_0^\dagger\subset S_{\mathcal{O}|\Omega}^*$ in the domains indicated. This automatically implies that the true relative modular operator extends the quadratic form $S_0^\dagger S_0$ defined on the domain above. What we mean is 
\begin{align*}
||\Delta_{\mathcal{O}|\Omega}^{1/2} a|\Omega\rangle||^2 = \langle\Omega|\mathcal{O}^*a a^{*}\mathcal{O}|\Omega\rangle = \langle\Omega| a^*\mathcal{O}\Delta_{\Omega}\mathcal{O}^*a|\Omega\rangle.
\end{align*} Of course to carry out perturbation theory, $\mathcal{O}\in\mathcal{A}_T$ as discussed previously.\\Similarly, with $\mathcal{O}\in\mathcal{A}$ and $\mathcal{O}^{-1}\in\mathcal{A}$, we have with $T_0 = (\mathcal{O}^{*})^{-1}\bar{S}_{\Omega}$ and $T_0^{\dagger} = S_{\Omega}^*\mathcal{O}^{-1}$, we get
\begin{align*}
T_0[a\mathcal{O}|\Omega\rangle] &= a^*|\Omega\rangle\\
T_0^\dagger[a'\mathcal{O}|\Omega\rangle] &= a^{'*}|\Omega\rangle,
\end{align*}and we get
\begin{align*}
||\Delta_{\Omega|\mathcal{O}}^{1/2} a\mathcal{O}|\Omega\rangle||^2 = \langle\Omega|a a^{*}|\Omega\rangle = \langle\Omega| \mathcal{O}^*a^*\Delta^{1/2}\mathcal{O}_J^{-1}\mathcal{O}_J^{*-1}\Delta_\Omega^{1/2}a\mathcal{O}|\Omega\rangle.
\end{align*}In principle, the same results occur if we restrict the operators $a\in\mathcal{A}_T\subset\mathcal{A}$ the Tomita algebra.

\subsection{Unbounded $\mathcal{O}$}
In the case when $\mathcal{O}$ is unbounded one needs to be careful. WLOG, we can choose $\mathcal{O} = \text{exp}(\beta\Phi)$ to be of exponential form. Then, the issue arises because vectors of the form $a|\Omega\rangle$ with $a\in\mathcal{A}$, are not in the domain of the local operators $\Phi$ while vectors of the form $a'|\Omega\rangle, a'\in\mathcal{A}'$ are (see Lemma 2.3 in \cite{Dreissler}). To navigate around this, one uses the Tomita algebra. We note
\begin{align*}
a_T|\Omega\rangle &= (a_T^*)^*|\Omega\rangle = J\Delta_{\Omega}^{1/2}(a_T^*)|\Omega\rangle = Ja_T^*(i/2)J|\Omega\rangle =b_T'|\Omega\rangle\\
b_T' &= Ja_T^*(i/2)J.
\end{align*}
Here $a_T(i/2)$ is the analytic continuiation of the modular evolved operator $a_T(t)$ evaluated at $t=i/2$. Formally, $a_T(i/2)=\Delta^{1/2}_\Omega a_T\Delta^{-1/2}_\Omega$. But it is more rigorous to think of $a_T(i/2)$ as simply the analytically continued operator $a_T(z)$ evaluated at $z=i/2$.

Now, $b_T'$ is a bounded operator in the commutant. Here we have used the fact that arbitrary complex powers of $\Delta_\Omega$ induce automorphisms of the Tomita algebra.

Then, for the case where $|\psi\rangle = \text{exp}(\beta\Phi)|\Omega\rangle$, we get with $X_0 = \bar{S}_{\Omega}\mathcal{O}$, $Y_0 = \mathcal{O}^{-1}\bar{S}_\Omega$, using the previous result,
\begin{align*}
X_0[a_T|\Omega\rangle] &= a_T^*|\psi\rangle,\\
Y_0[a_T|\psi\rangle]&=a_T^*|\Omega\rangle,
\end{align*}and the previous relations for the quadratic forms continue to hold here. 

\subsection{Domain of perturbation theory}
In our real time perturbation theory, for the definitions of $\delta$ to make sense, we need to work in the common domain of the operators $\Delta_0^{1/2}(\Delta+\beta)^{-1}$ and $\Delta_0^{-1/2}\Delta(\Delta+\beta)^{-1}$.

First lets consider the case where $\Delta=\Delta_{\Omega,\Psi}$, the relative modular operator between the vacuum and $|\Psi\rangle=\Psi|\Omega\rangle$, with $|\Psi\rangle$ invertible. $\Delta_0=\Delta_{\Omega}$, the modular operator of the vacuum. We choose $a\in\mathcal{A}_T(\Omega)$, the Tomita algebra of the vaccuum state.

Using the integral representation,
\begin{align*}
(\Delta+\beta)^{-1} a|\Psi\rangle&=\frac{i}{2}\text{lim}_{\epsilon\rightarrow 0}\int_{-\infty}^\infty\frac{dt}{\text{sinh}(\pi(t+i\epsilon))}\beta^{-it}\Delta^{it} a|\Psi\rangle\\&=\frac{i}{2}\text{lim}_{\epsilon\rightarrow 0}\int_{-\infty}^\infty\frac{dt}{\text{sinh}(\pi(t+i\epsilon))}\beta^{-it}\sigma^\Omega_t(a)\Delta^{it}|\Psi\rangle\\&=\frac{i}{2}\text{lim}_{\epsilon\rightarrow 0}\int_{-\infty}^\infty\frac{dt}{\text{sinh}(\pi(t+i\epsilon))}\beta^{-it}\sigma^\Omega_t(a)[D\Omega:D\Psi]_t|\Psi\rangle,
\end{align*}where $\sigma_t^\Omega(a)$ is the modular flow of $a$ and $[D\Omega:D\Psi]_t$ is the cocycle. Obviously, the $\epsilon\rightarrow 0$ limit exits on the RHS.

We observe that the integrand above is in the domain of $\Delta_0^{1/2}$. And, we get
\begin{align*}
\Delta_0^{1/2}\sigma^\Omega_t(a)[D\Omega:D\Psi]_t|\Psi\rangle = \sigma^\Omega_{t-i/2}(a)J\Psi^*[D\Psi:D\Omega]_t|\Omega\rangle\label{AAK}
\end{align*}where, by definition of the Tomita algebra, $\sigma_z^\Omega(a)$ is an analytic function of $z$. We have also used $[D\Omega:D\Psi]^*=[D\Psi:D\Omega]_t$. Moreover, the vector on the RHS seen as a function of $t$ is uniformly bounded. Therefore, for any $0<\epsilon<1$, the following integral exists,
\begin{align}
&\int_{-\infty}^\infty\frac{dt}{\text{sinh}(\pi(t+i\epsilon))}\beta^{-it}\sigma^\Omega_{t-i/2}(a)J\Psi^*[D\Psi:D\Omega]_t|\Omega\rangle.
\end{align}

Since $\Delta_0^{1/2}$ is a closed operator, Hille's theorem for the Bochner integral guarantees
\begin{align}
\int_{-\infty}^\infty\frac{dt}{\text{sinh}(\pi(t+i\epsilon))}\beta^{-it}\sigma^\Omega_t(a)[D\Omega:D\Psi]_t|\Psi\rangle\in D(\Delta_0^{1/2}).
\end{align}
and also
\begin{align}
&\Delta_0^{1/2}\int_{-\infty}^\infty\frac{dt}{\text{sinh}(\pi(t+i\epsilon))}\beta^{-it}\sigma^\Omega_t(a)[D\Omega:D\Psi]_t|\Psi\rangle \nonumber\\&= \int_{-\infty}^\infty\frac{dt}{\text{sinh}(\pi(t+i\epsilon))}\beta^{-it}\sigma^\Omega_{t-i/2}(a)J\Psi^*[D\Psi:D\Omega]_t|\Omega\rangle\nonumber\\&=\int_{-\infty}^\infty\frac{dt}{\text{sinh}(\pi(t+i\epsilon))}\beta^{-it}\sigma^\Omega_{t-i/2}(a)J\Psi^*\Delta_{\Psi,\Omega}^{it}|\Omega\rangle
\end{align}Since $|\Omega\rangle$ is in the domain of $\Delta_{\Psi,\Omega}^{1/2}$, the vector valued function $\Delta_{\Psi,\Omega}^{iz}|\Omega\rangle$ is analytic in the strip $\left\lbrace z:-1/2<\text{Im}(z)<0\right\rbrace$. This means $J\Psi^*\Delta_{\Psi,\Omega}^{i\bar{z}}|\Omega\rangle$ is analytic in the strip $\left\lbrace z:0<\text{Im}(z)<1/2\right\rbrace$ by the antilinearity of $J$. Therefore, $\sigma^\Omega_{z-i/2}(a)J\Psi^*\Delta_{\Psi,\Omega}^{i\bar{z}}|\Omega\rangle$ is analytic in this strip. Then, one can use the Cauchy theorem to write,
\begin{align*}
&\Delta_0^{1/2}\int_{-\infty}^\infty\frac{dt}{\text{sinh}(\pi(t+i\epsilon))}\beta^{-it}\sigma^\Omega_t(a)[D\Omega:D\Psi]_t|\Psi\rangle\\&=-i\sqrt{\beta}\int_{-\infty}^\infty\frac{dt}{\text{cosh}(\pi(t+i\epsilon))}\beta^{-it}\sigma^\Omega_{t}(a)J\Psi^*\Delta_{\Psi,\Omega}^{it+1/2}|\Omega\rangle.
\end{align*}This implies, by the dominated convergence theorem
\begin{align*}
&\text{lim}_{\epsilon\rightarrow 0}\Delta_0^{1/2}\biggl(\int_{-\infty}^\infty\frac{dt}{\text{sinh}(\pi(t+i\epsilon))}\beta^{-it}\sigma^\Omega_t(a)[D\Omega:D\Psi]_t|\Psi\rangle\biggr)\\&=-i\sqrt{\beta}\int_{-\infty}^\infty\frac{dt}{\text{cosh}(\pi t)}\beta^{-it}\sigma^\Omega_{t}(a)J\Psi^*\Delta_{\Psi,\Omega}^{it+1/2}|\Omega\rangle.
\end{align*}Obviously the last integral converges. Since $\Delta_0^{1/2}$ is a closed operator, this is sufficient to guarantee $a|\Psi\rangle$ is in the domain of $\Delta_0^{1/2}\Delta(\Delta+\beta)^{1/2}$. Next we look at
\begin{align*}
\Delta_0^{-1/2}\Delta(\Delta+\beta)^{1/2}a|\Psi\rangle
\end{align*}Again, an integral representation tells us
\begin{align*}
\Delta(\Delta+\beta)^{1/2}a|\Psi\rangle = -\frac{i}{2}\text{lim}_{\epsilon\rightarrow 0}\int_{-\infty}^\infty\frac{dt}{\text{sinh}(\pi(t-i\epsilon))}\beta^{-it}\sigma^\Omega_t(a)\Delta^{it}|\Psi\rangle.
\end{align*} Using an argument very similar to the prevous ones and using Cauchy's theorem we are able to write
\begin{align*}
\Delta(\Delta+\beta)^{1/2}a|\Psi\rangle = \frac{1}{2\sqrt{\beta}}\int_{-\infty}^\infty\frac{dt}{\text{cosh}(\pi t)}\beta^{-it}\sigma^\Omega_{t-i/2}(a)\Delta^{it}\Delta^{1/2}|\Psi\rangle.
\end{align*} Next, we note
\begin{align*}
\Delta^{1/2}|\Psi\rangle = J_{\Psi\Omega}|\Omega\rangle = \Theta'_{\Psi\Omega}|\Omega\rangle,
\end{align*} where $\Theta'_{\Psi\Omega}=J_{\Psi\Omega}J_\Omega$ is the conjugation cocycle which is a unitary in the commutant. This implies
\begin{align*}
\Delta(\Delta+\beta)^{1/2}a|\Psi\rangle = \frac{1}{2\sqrt{\beta}}\int_{-\infty}^\infty\frac{dt}{\text{cosh}(\pi t)}\beta^{-it}\sigma^\Omega_{t-i/2}(a)\Delta^{it}\Theta'_{\Psi\Omega}|\Omega\rangle.
\end{align*}

Finally, we can use
\begin{align*}
\Delta^{it}:= \Delta_{\Omega\Psi}^{it} = (\Delta_{\Psi\Omega}')^{-it}.
\end{align*}

Here, the prime on the last term indicates the relative modular operator with respect to the commutant. This implies
\begin{align*}
\Delta(\Delta+\beta)^{1/2}a|\Psi\rangle = \frac{1}{2\sqrt{\beta}}\int_{-\infty}^\infty\frac{dt}{\text{cosh}(\pi t)}\beta^{-it}\sigma^\Omega_{t-i/2}(a)[D\Psi:D\Omega]_{-t}'\sigma^\Omega_t(\Theta'_{\Psi\Omega})|\Omega\rangle,
\end{align*} where $[D\Psi:D\Omega]_{-t}'=(\Delta_{\Psi\Omega}')^{-it}(\Delta_\Omega')^{it}$ is the unitary cocycle in the commutant. Finally, we note that if $b$ is in the Tomita algebra and $|\alpha\rangle$ is in the domain of $\Delta_0^{-1/2}$, then $\Delta_0^{-1/2}b|\alpha\rangle=\sigma_{i/2}(b)\Delta^{-1/2}|\alpha\rangle$. This shows that the integrand in the previous equation is in the domain of $\Delta_0^{-1/2}$. Moreover, by Hille's theorem again,
\begin{align*}
&\Delta_0^{-1/2}\int_{-\infty}^\infty\frac{dt}{\text{cosh}(\pi t)}\beta^{-it}\sigma^\Omega_{t-i/2}(a)[D\Psi:D\Omega]_{-t}'\sigma^\Omega_t(\Theta'_{\Psi\Omega})|\Omega\rangle \\&= \frac{1}{2\sqrt{\beta}}\int_{-\infty}^\infty\frac{dt}{\text{cosh}(\pi t)}\beta^{-it}\sigma^\Omega_t(a)J\sigma^\Omega_t(\Theta'_{\Psi\Omega})^*([D\Psi:D\Omega]_{-t}')^*|\Omega\rangle
\end{align*}
we get that $\Delta(\Delta+\beta)^{1/2}a|\Psi\rangle$ is also in the domain of $\Delta_0^{-1/2}$ since the latter integral conveges.

\bibliographystyle{JHEP}

\bibliography{AQFT}
\end{document}

%% file: HLdefs.tex
\usepackage{color}

\newcommand{\nb}[1]{\color{blue}}

\newcommand{\hl}[1]{\color{magenta}}

\usepackage[normalem]{ulem}
\usepackage{amsmath}
\usepackage{enumerate}
\usepackage{amsfonts}
\usepackage{epsfig}
\usepackage{mathbbol}

\newcommand\half{{\ensuremath{\frac{1}{2}}}}

\newcommand\field[1]{{\ensuremath{\mathbb{{#1}}}}}

\newcommand\vev[1]{{\ensuremath{\left\langle{#1}\right\rangle}}}

\newcommand\ket[1]{\ensuremath{\lvert{#1}\rangle}}
\newcommand\bra[1]{\ensuremath{\langle{#1}\rvert}}

\newcommand{\CC}{\field{C}}

\newcommand{\bega}{\begin{gather}}
\newcommand{\eega}{\end{gather}}
\newcommand{\bi}{\begin{itemize}}
\newcommand{\ei}{\end{itemize}}
\newcommand{\ben}{\begin{enumerate}}
\newcommand{\een}{\end{enumerate}}
\newcommand{\bca}{\begin{cases}}
\newcommand{\eca}{\end{cases}}
\newcommand{\bln}{\begin{align}}
\newcommand{\eln}{\end{align}}
\newcommand{\bst}{\begin{split}}
\newcommand{\est}{\end{split}}
\def\ie{\begin{equation}\begin{aligned}}
\def\fe{\end{aligned}\end{equation}}
\newcommand{\bma}{\le(\begin{matrix}}
\newcommand{\ema}{\end{matrix}\ri)}

\newcommand\sig{\sigma}
\newcommand\Sig{\Sigma}
\newcommand\lam{\lambda}

\newcommand\om{\omega}
\newcommand\Om{\Omega}

\newcommand\de{{\ensuremath{{\delta}}}}
\newcommand\De{{\ensuremath{{\Delta}}}}

\newcommand\da{{\dagger}}

\newcommand\ov{\over}
\newcommand\ha{{\half}}

\def\le{\left}
\def\ri{\right}

\newcommand\sA{{\ensuremath{{\mathcal A}}}}
\newcommand\sB{{\ensuremath{{\mathcal B}}}}

\newcommand\sI{{\ensuremath{{\mathcal I}}}}

\newcommand\sH{{\ensuremath{{\mathcal H}}}}

\newcommand\sM{{\ensuremath{{\mathcal M}}}}

\newcommand\sO{{\ensuremath{{\mathcal O}}}}

%% file: flow_final_control.bbl
\providecommand{\href}[2]{#2}\begingroup\raggedright\begin{thebibliography}{10}

\bibitem{Witten:2018zxz}
E.~Witten, {\it {Notes on Some Entanglement Properties of Quantum Field
  Theory}},  \href{http://arxiv.org/abs/1803.0499}{{\tt arXiv:1803.0499}}.

\bibitem{Haag:1963dh}
R.~Haag and D.~Kastler, {\it {An Algebraic approach to quantum field theory}},
  {\em J. Math. Phys.} {\bf 5} (1964) 848--861.

\bibitem{Haag:1992hx}
R.~Haag, {\em {Local quantum physics: Fields, particles, algebras}}.
\newblock 1992.

\bibitem{Araki:1999ar}
H.~Araki, {\em {Mathematical theory of quantum fields}}.
\newblock 1999.

\bibitem{Araki:1975zu}
H.~Araki, {\it {Relative Entropy and Its Applications}},  in {\em {Proceedings,
  Les methodes mathematiques de la theorie quantique des champs: Marseille,
  France, June 23-27, 1975}}, pp.~61--79, 1976.

\bibitem{Araki:1976zv}
H.~Araki, {\it {Relative Entropy of States of Von Neumann Algebras}},  {\em
  Publ. Res. Inst. Math. Sci. Kyoto} {\bf 1976} (1976) 809--833.

\bibitem{vedral2002role}
V.~Vedral, {\it The role of relative entropy in quantum information theory},
  {\em Reviews of Modern Physics} {\bf 74} (2002), no.~1 197.

\bibitem{Borchers:2000pv}
H.~J. Borchers, {\it {On revolutionizing quantum field theory with Tomita's
  modular theory}},  {\em J. Math. Phys.} {\bf 41} (2000) 3604--3673.

\bibitem{Bousso:2015mna}
R.~Bousso, Z.~Fisher, S.~Leichenauer, and A.~C. Wall, {\it {Quantum focusing
  conjecture}},  {\em Phys. Rev.} {\bf D93} (2016), no.~6 064044,
  [\href{http://arxiv.org/abs/1506.0266}{{\tt arXiv:1506.0266}}].

\bibitem{Balakrishnan:2017bjg}
S.~Balakrishnan, T.~Faulkner, Z.~U. Khandker, and H.~Wang, {\it {A General
  Proof of the Quantum Null Energy Condition}},
  \href{http://arxiv.org/abs/1706.0943}{{\tt arXiv:1706.0943}}.

\bibitem{Casini:2008cr}
H.~Casini, {\it {Relative entropy and the Bekenstein bound}},  {\em Class.
  Quant. Grav.} {\bf 25} (2008) 205021,
  [\href{http://arxiv.org/abs/0804.2182}{{\tt arXiv:0804.2182}}].

\bibitem{Lashkari:2018nsl}
N.~Lashkari, {\it {Constraining Quantum Fields using Modular Theory}},
  \href{http://arxiv.org/abs/1810.0930}{{\tt arXiv:1810.0930}}.

\bibitem{Hislop:1981uh}
P.~D. Hislop and R.~Longo, {\it {Modular Structure of the Local Algebras
  Associated With the Free Massless Scalar Field Theory}},  {\em Commun. Math.
  Phys.} {\bf 84} (1982) 71.

\bibitem{Casini:2011kv}
H.~Casini, M.~Huerta, and R.~C. Myers, {\it {Towards a derivation of
  holographic entanglement entropy}},  {\em JHEP} {\bf 05} (2011) 036,
  [\href{http://arxiv.org/abs/1102.0440}{{\tt arXiv:1102.0440}}].

\bibitem{faulkner2014gravitation}
T.~Faulkner, M.~Guica, T.~Hartman, R.~C. Myers, and M.~Van~Raamsdonk, {\it
  Gravitation from entanglement in holographic cfts},  {\em Journal of High
  Energy Physics} {\bf 2014} (2014), no.~3 51.

\bibitem{lashkari2016gravitational}
N.~Lashkari, J.~Lin, H.~Ooguri, B.~Stoica, and M.~Van~Raamsdonk, {\it
  Gravitational positive energy theorems from information inequalities},  {\em
  Progress of Theoretical and Experimental Physics} {\bf 2016} (2016), no.~12.

\bibitem{Wall:2011hj}
A.~C. Wall, {\it {A proof of the generalized second law for rapidly changing
  fields and arbitrary horizon slices}},  {\em Phys. Rev.} {\bf D85} (2012)
  104049, [\href{http://arxiv.org/abs/1105.3445}{{\tt arXiv:1105.3445}}].
  [erratum: Phys. Rev.D87,no.6,069904(2013)].

\bibitem{Papadodimas:2013jku}
K.~Papadodimas and S.~Raju, {\it {State-Dependent Bulk-Boundary Maps and Black
  Hole Complementarity}},  {\em Phys. Rev.} {\bf D89} (2014), no.~8 086010,
  [\href{http://arxiv.org/abs/1310.6335}{{\tt arXiv:1310.6335}}].

\bibitem{Harlow:2016vwg}
D.~Harlow, {\it {The Ryu-Takayanagi Formula from Quantum Error Correction}},
  {\em Commun. Math. Phys.} {\bf 354} (2017), no.~3 865--912,
  [\href{http://arxiv.org/abs/1607.0390}{{\tt arXiv:1607.0390}}].

\bibitem{Casini:2017roe}
H.~Casini, E.~Teste, and G.~Torroba, {\it {Modular Hamiltonians on the null
  plane and the Markov property of the vacuum state}},  {\em J. Phys.} {\bf
  A50} (2017), no.~36 364001, [\href{http://arxiv.org/abs/1703.1065}{{\tt
  arXiv:1703.1065}}].

\bibitem{Harlow:2018tng}
D.~Harlow and H.~Ooguri, {\it {Symmetries in quantum field theory and quantum
  gravity}},  \href{http://arxiv.org/abs/1810.0533}{{\tt arXiv:1810.0533}}.

\bibitem{Rehren:1999jn}
K.-H. Rehren, {\it {Algebraic holography}},  {\em Annales Henri Poincare} {\bf
  1} (2000) 607--623, [\href{http://arxiv.org/abs/hep-th/9905179}{{\tt
  hep-th/9905179}}].

\bibitem{Jafferis:2015del}
D.~L. Jafferis, A.~Lewkowycz, J.~Maldacena, and S.~J. Suh, {\it {Relative
  entropy equals bulk relative entropy}},  {\em JHEP} {\bf 06} (2016) 004,
  [\href{http://arxiv.org/abs/1512.0643}{{\tt arXiv:1512.0643}}].

\bibitem{Faulkner:2017vdd}
T.~Faulkner and A.~Lewkowycz, {\it {Bulk locality from modular flow}},  {\em
  JHEP} {\bf 07} (2017) 151, [\href{http://arxiv.org/abs/1704.0546}{{\tt
  arXiv:1704.0546}}].

\bibitem{Faulkner:2018faa}
T.~Faulkner, M.~Li, and H.~Wang, {\it {A modular toolkit for bulk
  reconstruction}},  \href{http://arxiv.org/abs/1806.1056}{{\tt
  arXiv:1806.1056}}.

\bibitem{Bisognano:1976za}
J.~J. Bisognano and E.~H. Wichmann, {\it {On the Duality Condition for Quantum
  Fields}},  {\em J. Math. Phys.} {\bf 17} (1976) 303--321.

\bibitem{Borchers:1998}
H.~J. Borchers and J.~Yngvason, {\it {Modular groups of quantum fields in
  thermal states}},  {\em J. Math. Phys.} {\bf 40} (1999) 601--624,
  [\href{http://arxiv.org/abs/math-ph/9805013}{{\tt math-ph/9805013}}].

\bibitem{Saffary:2006}
T.~Saffary, {\it On the generator of massive modular groups},  {\em Letters in
  Mathematical Physics} {\bf 77} (2006), no.~3 235--248.

\bibitem{Longo:2009mn}
R.~Longo, P.~Martinetti, and K.-H. Rehren, {\it {Geometric modular action for
  disjoint intervals and boundary conformal field theory}},  {\em Rev. Math.
  Phys.} {\bf 22} (2010) 331--354, [\href{http://arxiv.org/abs/0912.1106}{{\tt
  arXiv:0912.1106}}].

\bibitem{Casini:2009}
H.~Casini and M.~Huerta, {\it {Reduced density matrix and internal dynamics for
  multicomponent regions}},  {\em Class. Quant. Grav.} {\bf 26} (2009) 185005,
  [\href{http://arxiv.org/abs/0903.5284}{{\tt arXiv:0903.5284}}].

\bibitem{Casini:2009sr}
H.~Casini and M.~Huerta, {\it {Entanglement entropy in free quantum field
  theory}},  {\em J. Phys.} {\bf A42} (2009) 504007,
  [\href{http://arxiv.org/abs/0905.2562}{{\tt arXiv:0905.2562}}].

\bibitem{Brunetti:2010}
R.~Brunetti and V.~Moretti, {\it {Modular dynamics in diamonds}},
  \href{http://arxiv.org/abs/1009.4990}{{\tt arXiv:1009.4990}}.

\bibitem{Tedesco:2014eaz}
G.~Tedesco, {\em {Modular structure of chiral Fermi fields in conformal quantum
  field theory}}.
\newblock PhD thesis, U. Gottingen (main), 2014.

\bibitem{Klich:2015}
I.~Klich, D.~Vaman, and G.~Wong, {\it {Entanglement Hamiltonians for chiral
  fermions with zero modes}},  {\em Phys. Rev. Lett.} {\bf 119} (2017), no.~12
  120401, [\href{http://arxiv.org/abs/1501.0048}{{\tt arXiv:1501.0048}}].

\bibitem{Cardy:2016}
J.~Cardy and E.~Tonni, {\it {Entanglement hamiltonians in two-dimensional
  conformal field theory}},  {\em J. Stat. Mech.} {\bf 1612} (2016), no.~12
  123103, [\href{http://arxiv.org/abs/1608.0128}{{\tt arXiv:1608.0128}}].

\bibitem{casini2017modular}
H.~Casini, E.~Test{\'e}, and G.~Torroba, {\it Modular hamiltonians on the null
  plane and the markov property of the vacuum state},  {\em Journal of Physics
  A: Mathematical and Theoretical} {\bf 50} (2017), no.~36 364001.

\bibitem{Klich:2017}
I.~Klich, D.~Vaman, and G.~Wong, {\it {Entanglement Hamiltonians and entropy in
  1+1D chiral fermion systems}},  {\em Phys. Rev.} {\bf B98} (2018) 035134,
  [\href{http://arxiv.org/abs/1704.0153}{{\tt arXiv:1704.0153}}].

\bibitem{lashkari2016modular}
N.~Lashkari, {\it Modular hamiltonian for excited states in conformal field
  theory},  {\em Physical review letters} {\bf 117} (2016), no.~4 041601.

\bibitem{Sarosi:2017rsq}
G.~Sarosi and T.~Ugajin, {\it {Modular Hamiltonians of excited states, OPE
  blocks and emergent bulk fields}},  {\em JHEP} {\bf 01} (2018) 012,
  [\href{http://arxiv.org/abs/1705.0148}{{\tt arXiv:1705.0148}}].

\bibitem{Leyland:1978iv}
P.~Leyland, J.~Roberts, and D.~Testard, {\it {DUALITY FOR QUANTUM FREE
  FIELDS}}, .

\bibitem{connes1973classification}
A.~Connes, {\it Une classification des facteurs de type iii},  in {\em Annales
  Scientifiques de l'{\'E}cole Normale Sup{\'e}rieure}, vol.~6, pp.~133--252,
  Elsevier, 1973.

\bibitem{connes1978homogeneity}
A.~Connes and E.~St{\o}rmer, {\it Homogeneity of the state space of factors of
  type iii1},  {\em Journal of Functional Analysis} {\bf 28} (1978), no.~2
  187--196.

\bibitem{dixmier:1971}
J.~Dixmier and O.~Mar{\' e}chal, {\it Vecteurs totalisateurs d'une algebre de
  von neumann},  {\em Comm. Math. Phys.} {\bf 22} (1971), no.~1 44--50.

\bibitem{robertson:1976}
G.~Robertson, {\it On the density of the invertible group in {$C^*$}-algebras},
   {\em Proceedings of the Edinburgh Mathematical Society} {\bf 20} (1976),
  no.~2 153--157.

\bibitem{YT}
Y.~Tanimoto, {\it Private communication}, .

\bibitem{Takesaki}
M.~Takesaki, {\em Theory of Operator algebras II}.
\newblock Springer-Verlag, 2003.

\bibitem{LLR}
N.~Lashkari, H.~Liu, and S.~Rajagopal, {\it {Perturbation Theory for the
  Logarithm of a Positive Operator}},
  \href{http://arxiv.org/abs/1811.0561}{{\tt arXiv:1811.0561}}.

\bibitem{Faulkner:2016mzt}
T.~Faulkner, R.~G. Leigh, O.~Parrikar, and H.~Wang, {\it {Modular Hamiltonians
  for Deformed Half-Spaces and the Averaged Null Energy Condition}},  {\em
  JHEP} {\bf 09} (2016) 038, [\href{http://arxiv.org/abs/1605.0807}{{\tt
  arXiv:1605.0807}}].

\bibitem{Headrick:2010zt}
M.~Headrick, {\it {Entanglement Renyi entropies in holographic theories}},
  {\em Phys. Rev.} {\bf D82} (2010) 126010,
  [\href{http://arxiv.org/abs/1006.0047}{{\tt arXiv:1006.0047}}].

\bibitem{penrose1971applications}
R.~Penrose, {\it Applications of negative dimensional tensors},  {\em
  Combinatorial mathematics and its applications} {\bf 1} (1971) 221--244.

\bibitem{wood2011tensor}
C.~J. Wood, J.~D. Biamonte, and D.~G. Cory, {\it Tensor networks and graphical
  calculus for open quantum systems},  {\em arXiv preprint arXiv:1111.6950}
  (2011).

\bibitem{bratteli81operator}
O.~Bratteli and D.~W. Robinson, {\em Operator Algebras and Quantum Statistical
  Mechanics: Volume 2: Equilibrium States models in Quantum Statistical
  Mechanics}.
\newblock Springer Science \& Business Media, 1981.

\bibitem{Dreissler}
W.~Dreissler, S.~J. Summers, and E.~H. Wichmann, {\it {On the connection
  between Quantum Fields and von Neumann algebras of Local Operators}},  {\em
  Commun. Math. Phys.} {\bf 105} (1986) 49--83.

\end{thebibliography}\endgroup
